\documentclass[a4paper,11pt]{article}
\pdfoutput=1 % if your are submitting a pdflatex (i.e. if you have
\usepackage{caption}
\usepackage{jcappub} % for details on the use of the package, please
\usepackage{aasmacros}              

\usepackage{lscape}
\usepackage{verbatim}
\usepackage{comment}
\usepackage{array}
\usepackage{subcaption}
\usepackage{multirow}
\usepackage{graphicx}
\graphicspath{{graphics/}}
\usepackage{amsmath}
\usepackage{amssymb}
\usepackage{rotating} 
\usepackage{xspace}
\usepackage{xcolor}
\usepackage{booktabs}
\usepackage[normalem]{ulem}
\usepackage{adjustbox}
\usepackage{lipsum}  
\newcommand{\Msun}{{\rm M}_\odot}

\newcommand{\be}{\begin{equation}}
\newcommand{\ee}{\end{equation}}
\newcommand{\nn}{\nonumber}

\title{The impact of the Large Magellanic Cloud on dark matter direct detection signals}

\author[a]{Adam Smith-Orlik,}
\author[b]{Nima Ronaghi,}
\author[b, c]{Nassim Bozorgnia,}
\author[d]{Marius Cautun,}
\author[e]{Azadeh Fattahi,}
\author[f]{Gurtina Besla,}
\author[e]{Carlos S. Frenk,}
\author[g]{Nicol\'{a}s Garavito-Camargo,}
\author[h, i]{Facundo A. G\'{o}mez,}
\author[j, k, l]{Robert J. J. Grand,}
\author[m]{Federico Marinacci,}
\author[n]{and Annika H. G. Peter}

\affiliation[a]{Department of Physics and Astronomy, York University,\\
4700 Keele Street, Toronto, Ontario M3J 1P3, Canada}
\affiliation[b]{Department of Physics, University of Alberta, Edmonton, Alberta T6G 2E1, Canada}
\affiliation[c]{Theoretical Physics Institute, University of Alberta,
Edmonton, Alberta T6G 2E1, Canada}
\affiliation[d]{Leiden Observatory, Leiden University, 
PO Box 9513, 2300 RA Leiden, the Netherlands}
\affiliation[e]{Institute for Computational Cosmology, Durham University,\\
South Road, Durham DH1 3LE, UK}
\affiliation[f]{Steward Observatory, University of Arizona,\\ 933 North Cherry Avenue,Tucson, AZ 85721, USA}
\affiliation[g]{Center for computational astrophysics, Flatiron Institute, \\
162 5th Ave, New York, NY 10010, USA}
\affiliation[h]{Departamento de F\'isica y Astronom\'ia, Universidad de La Serena,\\
Av. Juan Cisternas 1200 Norte, La Serena, Chile}
\affiliation[i]{Instituto de Investigaci\'on Multidisciplinar en Ciencia y Tecnolog\'ia, \\
Universidad de La Serena, Ra\'ul Bitr\'an 1305, La Serena, Chile}
\affiliation[j]{Instituto de Astrof\'isica de Canarias,\\ Calle Vía L\'actea s/n, E-38205 La Laguna, Tenerife, Spain}
\affiliation[k]{
Departamento de Astrof\'isica, Universidad de La Laguna, \\
Av.~del Astrof\'isico Francisco S\'anchez s/n, E-38206, La Laguna, Tenerife, Spain}
\affiliation[l]{Astrophysics Research Institute, Liverpool John Moores University,\\ 
146 Brownlow Hill, Liverpool, L3 5RF, UK}
\affiliation[m]{Department of Physics and Astronomy ``Augusto Righi'', University of Bologna,\\
via Gobetti 93/2, 40129 Bologna, Italy}
\affiliation[n]{CCAPP, Department of Physics, and Department of Astronomy,\\ 
The Ohio State University, Columbus, OH 43210 USA}

\abstract{
We study the effect of the Large Magellanic Cloud (LMC) on the dark matter (DM) distribution in the Solar neighborhood, utilizing the Auriga magneto-hydrodynamical simulations of Milky Way (MW) analogues that have an LMC-like system. We extract the local DM velocity distribution at different times during the orbit of the LMC  around the MW in the simulations. As found in previous idealized simulations of the MW-LMC system, we find that the DM particles in the Solar neighborhood originating from the LMC analogue  dominate  the high speed tail of the local DM speed distribution. Furthermore, the native  DM particles of the MW in the Solar region are boosted to higher speeds as a result of a response to the LMC's motion.
We simulate the signals expected in near future xenon, germanium, and silicon direct detection experiments, considering  DM interactions with target nuclei or electrons. We find that the presence of the LMC causes a considerable shift in the expected direct detection exclusion limits towards smaller cross sections and DM masses, with the effect being more prominent for low mass DM. Hence, our study shows, for the first time, that the LMC's influence on the local DM distribution is significant even in fully cosmological MW analogues. 
}

\begin{document}
\maketitle
\flushbottom

% BODY OF PAPER
\section{Introduction}
\label{Introduction}

Observational evidence points to the existence and abundance of dark matter (DM) in the Universe~\cite{Planck:2015fie}, and yet the  nature of DM remains unknown,  with the most popular theories suggesting that DM consists of one or more fundamental particle species. 
Direct detection searches aim to measure the small recoil energy of a target nucleus or electron  in an underground detector, after scattering with a massive DM particle. If DM consists of low mass axions instead, laboratory experiments can directly search for their conversion into photons in the detector. In order to interpret the results from these searches, knowledge of the phase-space distribution of DM in our Solar neighborhood is required. The most commonly adopted model for the DM halo of our galaxy is the Standard Halo Model (SHM)~\cite{Drukier:1986tm}. In the SHM, the DM particles are assumed to be distributed in an isothermal halo, and have an isotropic Maxwell-Boltzmann velocity distribution with a peak speed equal to the local circular speed.

Recent high resolution hydrodynamical simulations of galaxy formation find that while a Maxwellian velocity distribution models well the local DM velocity distribution of simulated Milky Way (MW) analogues, large halo-to-halo scatter exists in the distributions leading to large astrophysical uncertainties in the interpretation of direct detection results~\cite{Bozorgnia:2016ogo, Kelso:2016qqj, Sloane:2016kyi,  Bozorgnia:2017brl,  Bozorgnia:2019mjk, Poole-McKenzie:2020dbo, Lawrence:2022niq, Kuhlen:2013tra, Lacroix:2020lhn}.  
Hydrodynamical simulations also show that massive satellite mergers can  produce accreted stellar disks in some simulated galaxies, which may cause a degree of anisotropy in 
the local DM velocity distribution~\cite{Gomez:2017yzr}. The Galactic disk can also lead to the formation of a \emph{dark disk} component through accretion, with a surface density that has been  constrained using data from the \textit{Gaia} satellite~\cite{Widmark:2021gqx, Buch:2018qdr, Schutz:2017tfp}. Moreover, in light of data from 
Gaia~\cite{Gaia:2018ydn} and the Sloan Digital Sky Survey (SDSS)~\cite{SDSS:2000hjo} there is significant evidence that the MW contains kinematically distinct substructures due to its non-quiescent formation and merger history~\cite{Evans_2019,  Myeong:2019mvy, Helmi_2018, OHare:2019qxc, OHare:2018trr,  Aguado:2020kki, Necib:2018iwb, Necib:2019zbk} (see also \cite{Necib:2018igl, Bozorgnia:2018pfa, Maity:2022enp}). Recent hydrodynamical simulations and idealized models including specific substructures similar to those observed in Gaia show departures from the SHM that bear important implications for DM direct detection searches~\cite{Bozorgnia:2019mjk, Evans:2018bqy}.

In recent studies~\cite{Besla:2019xbx, Donaldson:2021byu, GaravitoCamargo:2021tcp, Garavito-Camargo:2020lqm, Conroy_2021, Petersen_2020,  Peterson_2020_MNRAS, Cunningham:2020nlo}, special attention has been paid to the effect of the Large Magellanic Cloud (LMC) on the local DM  distribution and the DM halo of the MW. Using idealized N-body simulations to fit the kinematics of the MW-LMC system, ref.~\cite{Besla:2019xbx} found that the high speed tail  of the DM velocity distribution in the Solar neighborhood is impacted both by DM particles that originated from the LMC and by native DM particles of the MW whose orbits have been altered considerably due to the gravitational pull of the LMC. 
Similarly, ref.~\cite{Donaldson:2021byu} used idealized models of the MW-LMC system and showed that close pericenter passage of the LMC  
results in boosts in the DM velocity distribution in the Solar region with the DM particles of the MW also being boosted by the reflex motion caused by the LMC at infall~\cite{Gomez:2014tda}, consistent with the results of ref.~\cite{Besla:2019xbx}.

Idealized simulations, such as those studied in refs.~\cite{Donaldson:2021byu, Besla:2019xbx}, can match the exact orbit and properties of the LMC in the MW halo. However, it remains to be determined that their findings are valid for fully cosmological halos with multiple accretion events over their formation history. 
In particular, an important question is whether a recent ($ \lesssim 100$~Myr) and close ($\lesssim 100$~kpc) pericentric approach of a massive satellite can significantly impact  the local DM distribution, despite the varied assembly history of a MW analogue in a fully cosmological setup. 
Another relevant question is whether the boost in the local DM velocity distribution is a generic feature for any Sun-LMC geometry, or if there are  particular geometries that augment this effect. Cosmological simulations that sample potential MW formation histories are,  therefore, necessary to characterize  the extent of the signatures of the MW-LMC interaction, 
and can provide further crucial insight on the LMC's effect, as well as the halo-to-halo uncertainties in the results~\cite{Santos-Santos:2021yal}.

In this paper, we use the Auriga cosmological magneto-hydrodynamical simulations~\cite{Grand:2016mgo} to study the effect of LMC-like systems on the local DM distribution of the host MW-like galaxies and their implications for DM direct detection.
The paper is structured as follows. In section~\ref{sec: simulations} we discuss the simulations details, our selection criteria for choosing MW-LMC analogues (section~\ref{sec: selection criteria}), and how we specify the Sun's position in the simulations (section~\ref{sec: Sun-LMC Geometry}). 
In sections~\ref{sec: DM density} and \ref{sec: velocity distributions}, we present the local DM density and velocity distributions extracted from the simulations, respectively. In section~\ref{sec: halo integrals}, we discuss the analysis of the so-called halo integral, which is an important input in DM direct detection computations, and show how the LMC impacts it. 
In section~\ref{sec: direct detection}, we discuss the implications of the LMC for DM direct detection signals, considering both DM-nucleus (section~\ref{sec: nuclear}) and DM-electron (section~\ref{sec: electron}) scattering. Finally, we conclude with a brief discussion and conclusion in section~\ref{sec: discussion}.

\section{Simulations}
\label{sec: simulations}

In this work we use the magneto-hydrodynamical simulations of MW mass halos from the Auriga project~\citep{Grand:2016mgo}. The original Auriga simulation suite includes 30 cosmological zoom-in simulations of isolated MW-mass halos, selected from a $100^3$~Mpc$^3$ periodic cube (L100N1504) from the EAGLE project~\cite{Schaye2015,Crain2015}. The simulations were performed using the moving-mesh code Arepo~\citep{Springel2010} and use a galaxy formation subgrid model which includes metal cooling, black hole formation, AGN and supernova feedback, star formation, and background UV/X-ray photoionisation radiation~\cite{Grand:2016mgo}. Planck-2015~\citep{Planck:2015fie} cosmological parameters are used for the simulations: $\Omega_{m}=0.307$, $\Omega_{\rm bar}=0.048$, $H_0=67.77~{\rm km~s^{-1}~Mpc^{-1}}$. We use the standard resolution level (Level 4) of the simulations with DM particle mass, $m_{\rm DM} \sim 3\times 10^5~\Msun$, baryonic mass element, $m_b=5\times10^4~\Msun$, and the Plummer equivalent gravitational softening, $\epsilon=370$~pc~\citep{Power:2002sw,Jenkins2013}. The Auriga simulations reproduce the observed stellar masses, sizes, rotation curves, star formation rates and metallicities of present day MW-mass galaxies.

\subsection{Selection criteria for MW-LMC analogues}
\label{sec: selection criteria}

To study the effect of the LMC on the local DM distribution, we first need to select simulated LMC analogues that have properties similar to the observed LMC. The LMC has just passed its first pericenter approach $\sim 50$~Myr ago~\cite{Besla:2007kf}.  
We will therefore use the properties of the LMC, at or close to its first pericenter passage. 
The present day stellar mass of the LMC from observations is $\sim 2.7 \times 10^9~\Msun$~\cite{vanderMarel:2002kq}, the LMC's first pericenter distance was at $\sim 48$~kpc~\cite{Besla:2007kf}, and its speed at pericenter with respect to the MW was $340 \pm 19$~km/s~\cite{Salem_2015}. The current speed of the LMC with respect to the MW's center is $321 \pm 24$~km/s~\cite{Kallivayalil:2013xb}.

The large phase-space of potential MW-LMC interactions makes it difficult to find an exact analogue in cosmological simulations, especially when we are dealing with only 30 MW-mass halos. To improve these chances, we not only consider present day matches, but follow back in time the history of the simulated MW analogues to find if a MW-LMC like interaction took place since redshift $z=1$ (i.e.~within the last 8 Gyrs). 
From the 30 Auriga halos, we first identify those that have an LMC analogue using the following criteria: (i) stellar mass\footnote{The stellar mass of the LMC analogue is the mass of all the stars associated with the LMC-like satellite as identified by the SUBFIND algorithm~\cite{Springel:2000qu}.} of the LMC analogue is $>5 \times 10^8~\Msun$, and (ii) distance of the LMC analogue from host at first pericenter is in the range of $[40, 60]$~kpc. With these criteria, we identify 15 MW-LMC analogues, which we study at the simulation snapshot (i.e.~output in time) closest to the LMC's first pericenter approach. We consider this snapshot as a proxy for the present day MW-LMC system. Notice that the average time between the simulation snapshots is $\sim 150$~Myr, so it is difficult to precisely obtain the present day snapshot for the 15 MW-LMC analogues. This large snapshot spacing is a limitation of the cosmological simulation approach, and we discuss below how we address this limitation.

In table~\ref{tab:peri}, we list some of the properties of the 15 MW-LMC analogues. The first two columns of the table show the halo ID of the MW-LMC analogues and the corresponding Auriga ID of the MW halos hosting the LMC. The next five columns list the properties of the analogues at the snapshot closest to LMC's first pericenter approach. From left to right, these include the distance of the LMC analogues from host, $r_{\rm LMC}$, the lookback time, $t_{\rm LB}$, the stellar mass of the MW  analogues, $M_{\ast}^{\rm MW}$, the virial mass\footnote{Virial mass is defined here as the mass enclosed within a spherical radius where the mean enclosed matter density is 200 times the critical density of the Universe.}  of the MW analogues, $M_{200}^{\rm MW}$, and the stellar mass of the LMC analogues, $M_{\ast}^{\rm LMC}$. The last column  lists the virial mass of the LMC analogues at infall, $M_{\rm Infall}^{\rm LMC}$. The speed of the LMC analogues with respect to the center of the MW analogues is in the range of $[205, 376]$~km/s at the snapshot closest to first pericenter approach.

\begin{table}
\centering
\begin{tabular}{|c|c|c|c|c|c|c|c|}
\hline
\multirow{2}{*}{Halo ID} & \multirow{2}{*}{Auriga ID} & $r_{\rm LMC}$ & $t_{\rm LB}$ & $M_\ast^{\rm MW}$ & $M_{200}^{\rm MW}$ & {$M_\ast^{\rm LMC}$} & $M^{\rm LMC}_{\rm Infall}$  \, \\
 &  & [kpc] & [Gyr]  & [$10^{9}~\Msun$] & [$10^{11}~\Msun$] & [$10^9~\Msun$] & $[10^{11}~\Msun]$\ \\ 
 \hline
       1 & Au-1 & 53.1 & 5.1 & $15$ & $4.0$ & $0.66$ & 0.31 \\
       2 & Au-7 & 49.2 & 4.2 & $23$ & $9.3$ & $0.56$ & 0.31\\
       3 & Au-12 & 49.4 & 4.6 & $33$ & $11$ & $0.79$ & 0.34 \\
       4 & Au-13 & 45.8 & 6.7 & $26$ & $9.5$ & $2.4$ & 0.82 \\
       5  & Au-13 & 56.7 & 7.4 & $16$ & $7.2$ & $3.1$ & 1.8\\
       6 & Au-14 & 45.6 & 6.7  & $37$ & $13$ & $3.3$ & 1.1\\
       7 & Au-14 & 49.9 & 2.3 & $93$ & $16$ & $0.99$ & 0.32\\
       8 & Au-17 & 54.0 & 7.1 & $50$ & $8.9$ & $0.85$ & 0.36\\
       9 & Au-19 & 40.9 & 6.2 & $18$ & $6.6$ & $1.6$ & 0.73\\
       10 & Au-19 & 50.8 & 5.4 & $21$ & $12$ & $9.1$ & 3.3\\
       11 & Au-21 & 55.5 & 3.3 & $67$ & $17$ & $4.8$ & 1.5\\
       12 & Au-23 & 41.0 & 5.9 & $60$ & $16$ & $2.5$ & 1.4\\
       13 & Au-25 & 43.2 & 1.0 & $37$ & $12$ & $15$ & 3.2\\
       14 & Au-27 & 58.9 & 6.3 & $56$ & $16$ & $1.0$ & 0.84\\
       15 & Au-30 & 56.0 & 6.3 & $73$ & $9.7$ & $2.5$ & 1.2\\
\hline
\end{tabular}
\caption{\label{tab:peri}Properties of the 15 MW-LMC analogues. The first two columns list the halo ID and Auriga ID of the analogues. The 3rd-7th columns list the properties of the analogues at the simulation snapshot closest to LMC's first pericenter approach: distance of the LMC analogues from host, $r_{\rm LMC}$, lookback time, $t_{\rm LB}$, stellar mass of the MW  analogues, $M_{\ast}^{\rm MW}$, virial mass of the MW analogues, $M_{200}^{\rm MW}$, and the stellar mass of the LMC analogues, $M_{\ast}^{\rm LMC}$. The last column lists the LMC's virial mass at infall, $M^{\rm LMC}_{\rm Infall}$.}
\end{table}

The halo mass of the actual LMC at infall is estimated to be $\sim(1-3) \times 10^{11}~\Msun$~\cite{Moster:2012fv, Laporte:2016vuu, Penarrubia:2015hqa, Erkal:2019}. As it can be seen from the last column of table~\ref{tab:peri}, five of our selected LMC analogues have $M_{\rm Infall}^{\rm LMC} \lesssim 0.4 \times 10^{11}~\Msun$. A related parameter of interest  is the ratio of the halo mass of LMC at infall to the MW halo mass. For five of the MW-LMC analogues, $M_{\rm Infall}^{\rm LMC}/M_{200}^{\rm MW}\lesssim 0.05$, which is about 3 times smaller than the LMC to MW mass ratio estimate from observations. These LMC analogues may have a smaller overall impact on their host halos, contribute less DM particles in the Solar neighborhood, and cause a less significant reflex motion~\cite{Gomez:2014tda, Peterson_2020_MNRAS, Donaldson:2021byu} in the MW analogues. However, we note that 
it is difficult to directly compare the halo masses of the LMC analogues from cosmological simulations with estimates from earlier studies based on observations, since those typically assume fixed mass in time or even a point mass. We therefore include the LMC analogues with the smaller halo mass at infall in our study to increase our sample size. In section~\ref{sec: DM density}, we discuss the implications of the smaller LMC to MW mass ratio for the number of DM particles from the LMC in our local neighborhood.

To study in more detail how the LMC affects the local DM distribution at different times in its orbit, we  select  one MW-LMC analogue, halo 13 corresponding to the Auriga 25 halo (hereafter Au-25) and its LMC analogue, for further study. This system has the second largest LMC halo mass at infall, close to the upper limit of the range estimated from observations. As a consequence, it also has a large $M_{\rm Infall}^{\rm LMC}/M_{200}^{\rm MW}=0.27$. We rerun the simulation for halo 13 with finer snapshots close to the LMC's pericenter approach. The average time between snapshots near pericenter in this new run is $\sim 10$~Myr. We consider four representative snapshots for halo 13: 
\emph{Iso.}~is the snapshot which takes place when the MW and the LMC analogues are maximally separated (i.e.~first apocenter before infall) at $\sim 2.83$~Gyr before the present day snapshot, acting as our proxy for an isolated MW; \emph{Peri.}~is the simulation equivalent of the point of closest approach (pericenter) of the LMC at $\sim 133$~Myr before the present day snapshot; \emph{Pres.}~is the closest snapshot to the present day separation of the observed MW and LMC system; and \emph{Fut.}~is a proxy for the MW-LMC system at a future point in time, $\sim 175$~Myr after the present day snapshot. 
 
 In table~\ref{tab:snapshots}, we summarize the description of these four snapshots, specify their times relative to the present day snapshot, and list the distance of the LMC analogue from host at each snapshot. The distance and speed of the LMC analogue with respect to its host at the present day snapshot are $\sim 50$~kpc and 317~km/s, respectively, which are remarkably close matches to the observed values\footnote{Notice that the distance of the LMC analogue at its pericentre approach is smaller than the value inferred from observations.}. Notice that when we refer to the ``present day'' snapshot for the re-simulated halo 13 throughout this work, we are referring to the \emph{Pres.}~snapshot. 

   \begin{table}[t]
    \centering
    \begin{tabular}{|c|c|c|c|}
      \hline
       Snapshot & Description & $t-t_{\rm Pres.}$~[Gyr] & $r_{\rm LMC}$~[kpc] \\
       \hline
       Iso. & Isolated MW analogue & $-2.83$ & 384  \\
       Peri. & LMC's first pericenter approach & $-0.133$ & 32.9\\
       Pres.  & Present day MW-LMC analogue & 0 & 50.6 \\
       Fut. & Future MW-LMC analogue & 0.175 & 80.3\\
      \hline
    \end{tabular}
\caption{Description of the four representative snapshots in halo 13, their times relative to the present day snapshot, and the distance of the LMC analogue from host at each snapshot.
}
\label{tab:snapshots}
\end{table}

In the rest of this paper we present some general results for the 15 selected MW-LMC analogues at the snapshot closest to the LMC's first pericenter approach, and then focus on halo 13 to study how the LMC impacts the local DM distribution during its orbit around the MW.

\subsection{Matching the Sun-LMC geometry}
\label{sec: Sun-LMC Geometry}

The geometry of the observed Sun-LMC system is such that the LMC is predominantly moving in the opposite direction of the Solar motion. This leads to large relative speeds of the particles originating from the LMC with respect to the Sun, and results in a boost in the DM velocity distribution in the Solar region~\cite{Besla:2019xbx}. Ref.~\cite{Besla:2019xbx} showed that matching the Sun-LMC geometry in their idealized simulations to the observed geometry is crucial for an accurate understanding of LMC's impact on the local DM distribution.

In the simulations, the position of the Sun is not specified a priori and the LMC analogues have different phase-space coordinates compared to the real MW-LMC system. Therefore, we need to choose a position for the Sun in each MW analogue based on a set of criteria for obtaining a match to the observed Sun-LMC geometry. We would also like to explore to what extent it is critical to match the exact Sun-LMC geometry in the simulations in order to have a significant effect on the local DM velocity distribution. In this section, we first discuss our procedure for obtaining all possible positions for the Sun in the simulations that approximately match the Sun-LMC geometry in observations. We next discuss how we specify the ``best fit'' Sun's position in the simulations that provides the best match to the observed Sun-LMC geometry.

\begin{figure}[t]
    \centering
    \includegraphics[scale=0.2]{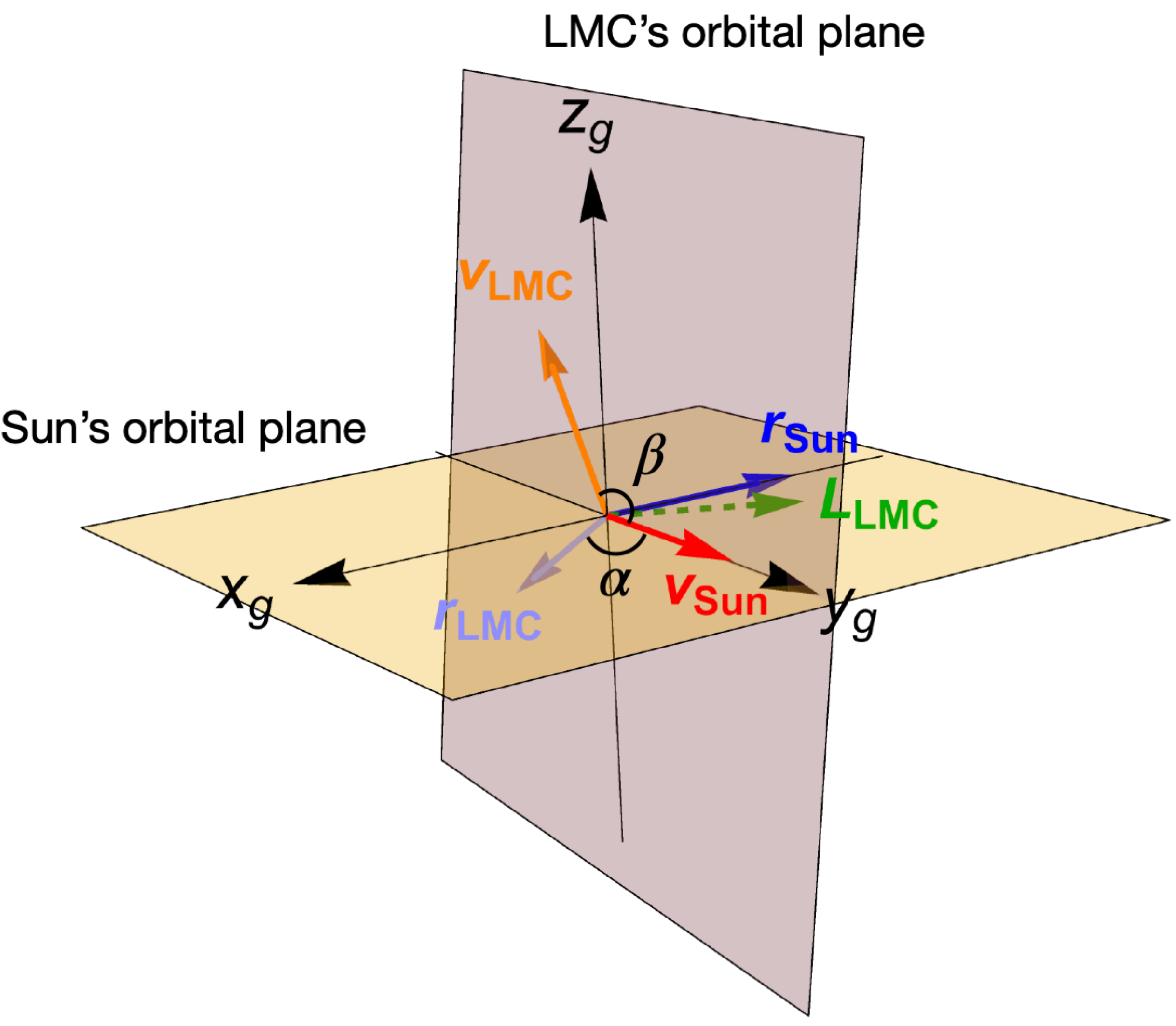}
    \caption{Diagram showing  the observed Sun-LMC geometry. The blue and red vectors specify the directions of the position, ${\bf r}_{\rm Sun}$, and velocity, ${\bf v}_{\rm Sun}$, of the Sun, while the light blue and orange vectors specify the directions of the position, ${\bf r}_{\rm LMC}$, and velocity, ${\bf v}_{\rm LMC}$, of the LMC. The angle $\alpha$, between ${\bf r}_{\rm LMC}$ and ${\bf v}_{\rm Sun}$, and the angle $\beta$, between ${\bf v}_{\rm LMC}$ and ${\bf v}_{\rm Sun}$ are specified. The dashed green vector shows the direction of the orbital angular momentum of the LMC, ${\bf L}_{\rm LMC}$. The orbital planes of the Sun and the LMC, which are nearly perpendicular, are also shown. 
    }
    \label{fig: geometry}
\end{figure}

Figure~\ref{fig: geometry} shows the observed geometry of the Sun-LMC system in the Galactocentric reference frame defined in the following way. The origin of the reference frame is on the Galactic center, the $x_g$ and $y_g$ axes are aligned with the Sun's orbital  plane, the $x_g$-axis points from the Sun towards the Galactic center, the $y_g$-axis is in the direction of the Galactic rotation, and the $z_g$-axis is towards the North Galactic Pole. The directions of the Sun's position, ${\bf r}_{\rm Sun}$, Sun's velocity, ${\bf v}_{\rm Sun}$, LMC's position, ${\bf r}_{\rm LMC}$, LMC's velocity, ${\bf v}_{\rm LMC}$, and the orbital angular momentum of the LMC, ${\bf L}_{\rm LMC}$, are specified in the diagram.

 In the simulations, we define the center of the MW and LMC analogues to be the position of the particle (star, gas, DM, or black hole) in each halo that has the lowest gravitational potential energy. The velocity of the MW and LMC analogues in the simulation reference frame is defined as  the centre of mass velocity of all bound particles to each halo, obtained using the SUBFIND algorithm~\cite{Springel:2000qu}. The position and velocity of the LMC analogue are then found with respect to the center of the MW analogue.

To find the possible positions for the Sun in the simulations that match the observed Sun-LMC geometry, we could impose a set of constraints on the angular coordinates of both the position and velocity vectors of the LMC analogues as seen from the Solar position in the simulation. However, the position and velocity vectors of the LMC analogues can change rapidly when the satellite is close to its pericentric approach. Thus, a better criterion for finding the Sun's position and the orientation of its orbital plane in the simulations is to ensure that the orbital plane of the LMC analogue makes the same angle with the Sun's orbital plane  as in observations.

We therefore proceed as follows to match the observed Sun-LMC geometry in the simulations. First, we choose a stellar disk orientation by requiring that the angle between the  angular momentum of the stellar disk and the orbital angular momentum of the LMC analogue, ${\bf L}_{\rm LMC}^{\rm sim}$, is the same as the observed MW-LMC pair. As seen in figure~\ref{fig: geometry}, the LMC's orbital angular momentum inferred from observations is nearly perpendicular to the angular momentum of the stellar disk. Hence, we can vary the latter on nearly  a full circle, resulting in multiple allowed stellar disk orientations for the simulated MW analogue. In particular, given the direction of ${\bf L}_{\rm LMC}^{\rm sim}$, we numerically solve for the direction of the disk's angular momentum by varying one of its angular coordinates every $10^\circ$, and finding the other angular coordinate such that it matches the observed MW-LMC orientation. Due to this sampling, the number of the allowed disk orientations we find varies from $\sim 20$ to over 30, depending on the MW-LMC analogue. Notice that these disk orientations are not necessarily aligned with the actual stellar disk of the MW analogue, but we consider them since they match the observed MW-LMC geometry, which is important for our study. 

Previous studies using the EAGLE and APOSTLE simulations show that the stellar disk does not have a significant effect on the local DM velocity distribution~\cite{Bozorgnia:2016ogo, Schaller:2016uot}. However, using idealized simulations refs.~\cite{Petersen:2016vck, Donaldson:2021byu} find that the presence of the stellar disk and its non-axisymmetric evolution  lead to secular processes, which  
can boost the local DM velocity distribution. 
We note that a number of Auriga halos have a small  DM component rotating with the stellar disk due to accretion events~\cite{Gomez:2017yzr}, but those halos are not part of our MW-LMC analogue sample.

In the next step, we find the Sun's position with respect to the center of the MW analogue for each allowed disk orientation by requiring that the angles between the LMC's orbital angular momentum and the Sun's position and velocity vectors are as close as possible to the observed values. From these first two steps, we obtain the position and velocity vectors of the Sun for each allowed disk orientation. Therefore, for each halo we obtain multiple allowed positions for the Sun, due to the multiple allowed disk orientations. In section~\ref{sec: halo integrals}, we will study how the MW-LMC interaction signatures vary depending on these  Sun’s positions.

We next proceed to find the best fit Sun's position. As seen in figure~\ref{fig: geometry}, 
the Sun's position vector is nearly along the same direction as the angular momentum of the LMC, and therefore varies only slightly for different disk orientations. On the other hand, the Sun's velocity vector varies on nearly a full circle, requiring further matching to observations. We define the cosine angles, 
\begin{align}
   \cos\alpha &\equiv \hat{\textbf{v}}_{\rm Sun}^{\rm sim} \cdot \hat{\textbf{r}}_{\rm LMC}^{\rm sim}\; , \nn \\ 
    \cos\beta &\equiv  \hat{\textbf{v}}_{\rm Sun}^{\rm sim} \cdot \hat{\textbf{v}}_{\rm LMC}^{\rm sim} \; ,
    \label{eq: cosine angles}
\end{align}
where $\hat{\textbf{v}}_{\rm Sun}^{\rm sim}$ is in the direction of the velocity of the Sun with respect to center of the MW analogue, while $\hat{\textbf{r}}_{\rm LMC}^{\rm sim}$ and $\hat{\textbf{v}}_{\rm LMC}^{\rm sim}$ are in the directions of the position and velocity vectors of the LMC analogue with respect to the center of the MW analogue. 
In the last step, we select the orientation that leads to the closest match with the observed values for the cosine angles,
\begin{align}
    \hat{\textbf{v}}_{\rm Sun}^{\rm obs} \cdot \hat{\textbf{r}}_{\rm LMC}^{\rm obs} &= -0.835\; , \nn\\
    \hat{\textbf{v}}_{\rm Sun}^{\rm obs} \cdot \hat{\textbf{v}}_{\rm LMC}^{\rm obs} &= -0.709 \, . 
    \label{eq: best fit cosine angles}
\end{align}
The best fit Sun's velocity vector in the simulations is found by choosing the values of $\cos\alpha$ and $\cos\beta$  that minimize the sum of the squared differences with the values obtained from observations, given in eq.~\eqref{eq: best fit cosine angles}. This, in turn, determines the best fit Sun's position.

\section{Local dark matter distribution}
\label{sec: distributions}

Computations of DM direct detection event rates strongly depend on the assumptions made for the DM  distribution in the Solar neighborhood. In sections \ref{sec: DM density} and \ref{sec: velocity distributions}, we present the DM density and velocity distribution in the Solar neighborhood extracted from the simulated MW-LMC analogues, and discuss the effect of the LMC on the results.

For each possible Sun's position (and velocity) which matches the observed Sun-LMC geometry, we consequently have the orientation of the $(x_g, y_g, z_g)$ axes of the Galactic reference frame defined in section~\ref{sec: Sun-LMC Geometry}. We then transform the positions and velocities of the simulation particles to  this Galactic reference frame.  

\begin{figure}[t]
    \centering
    \includegraphics[scale=0.45]{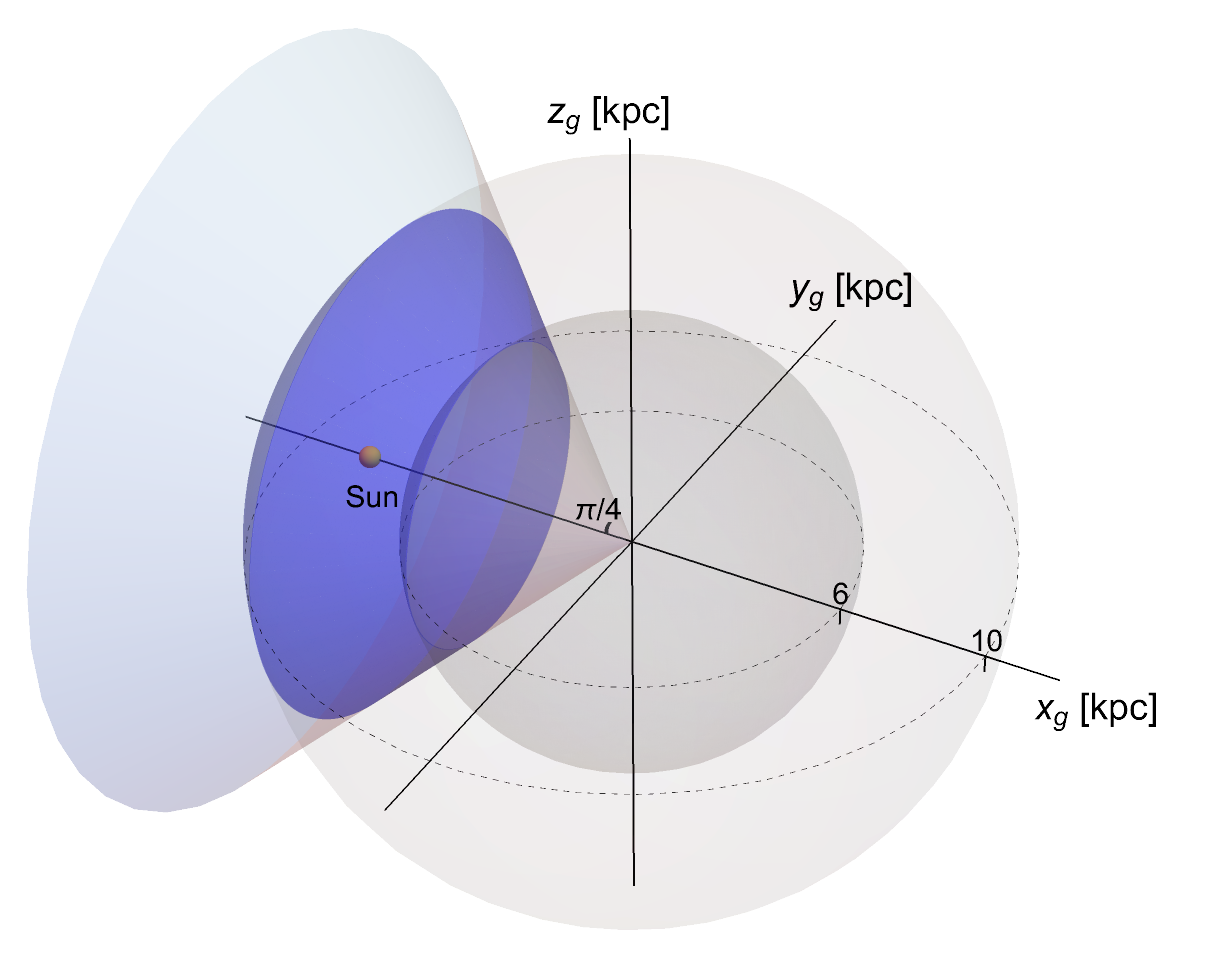}
    \caption{Diagram of the Solar region (blue) chosen as the overlap of the volume of a cone projected from the Galactic center with an opening angle of $\pi/4$ radians and its axis fixed on the position of the Sun, with the volume enclosed between two concentric spheres with radii of 6 and 10~kpc from the Galactic center. For illustration, the Sun is placed on the cone's axis at a Galactocentric distance of 8 kpc.
    }
    \label{fig: solar region}
\end{figure}

To define the \emph{Solar region}, with the Sun at a galactocentric distance of $\sim 8$~kpc, we first consider the region enclosed within a spherical shell between 6 to 10 kpc from the Galactic center of the MW analogue. We then consider a cone with an opening angle of $\pi/4$ radians, its vertex at the Galactic center, and its axis aligned with the position of the Sun as obtained from the procedure discussed in section~\ref{sec: Sun-LMC Geometry}. The overlap of the spherical shell and the cone constitutes the Solar region, shown as the shaded blue region in figure~\ref{fig: solar region}. The size of the Solar region is chosen to be large enough to include several thousand DM particles, 
and small enough to retain sensitivity to the best fit Sun's position. In sections~\ref{sec: DM density} and \ref{sec: Variation due to Sun-LMC Geometry}, we discuss the impact of changing the size of the Solar region on the local DM density, the percentage of the DM particles originating from the LMC in the Solar region, and the high speed tails of the halo integrals.

Since the set of allowed and best fit Sun's positions we find using the procedure described in section~\ref{sec: Sun-LMC Geometry} vary for each halo and snapshot, the Solar region is  different for each MW analogue and each snapshot.

\begin{table}[t!]
    \centering
    \begin{tabular}{|c|c|c|c|c|c|c|}
      \hline
      Halo ID  & $N_{\rm MW}$ & $N_{\rm LMC}$ & $\rho_{\chi}$~[GeV$/$cm$^3$] & $\kappa_{\rm LMC}$~[\%] & $\kappa_{\rm LMC}$ Range~[\%] & $v_{\rm esc}^{\rm det}$~[km/s] \\
       \hline
       1  & 7,760 & 11 & 0.21 & 0.14 & $[0.14-0.21]$ & 651  \\
       2  & 8,581 & 55 & 0.23 & 0.64 & $[0.53-0.65]$ & 720 \\
       3 & 11,621 & 3 & 0.35 & 0.026 & $[0.025-0.028]$ & 714 \\
       4  & 12,483 & 12 & 0.34 & 0.096 & $[0.088-0.12]$ & 737 \\
       5  & 8,669 & 136 & 0.24 & 1.5 & $[1.4-1.6]$ & 707 \\
       6  & 13,290 & 5 &  0.38 & 0.038 & $[0.029-0.046]$ &  734 \\
       7  & 18,467 & 6 &  0.53 & 0.032 & $[0.032-0.034]$ & 766 \\
       8  & 12,949 & 1 & 0.38 & 0.0077 & $[0.0077-0.0082]$ & 712 \\
       9  & 11,892 & 12 & 0.36 & 0.10 & $[0.069-0.13]$ & 715 \\
       10 & 12,405 & 361 & 0.39 & 2.8 & $[2.8-3.1]$ & 791\\
       11  & 14,132 & 4 & 0.43 & 0.028 & $[0.021-0.039]$ & 758 \\
       12  & 16,427 & 28 & 0.53 & 0.17 & $[0.17-0.21]$ & 783 \\
       13 & 10,814 & 254 &  0.34 & 2.3 & $[2.3-3.0]$ & 831 \\
       14  & 20,001 & 52 & 0.60 & 0.26 & $[0.26-0.31]$ & 776 \\
       15 & 10,641 & 128 &  0.32 & 1.2 & $[0.81-1.3]$ & 819 \\
      \hline
    \end{tabular}%}
\caption{Various quantities for the 15 MW-LMC analogues in the Solar region, at the simulation snapshot closest to the LMC's pericenter approach: halo ID, the number of native DM particles of the MW, $N_{\rm MW}$, the number of DM particles originating from the LMC, $N_{\rm LMC}$, the local DM density, $\rho_\chi$, the  percentage of the DM particles originating from the LMC in the Solar region, $\kappa_{\rm LMC}$, for the best fit Sun's position, the range that $\kappa_{\rm LMC}$ can span across the different allowed Sun's positions, and the local escape speed from the MW in the detector rest frame, $v_{\rm esc}^{\rm det}$. All columns, except the 6th, list the quantities for the best fit Sun's position.} 
    \label{tab:Localrho}
  \end{table}  

The number of the native DM particles of the MW, $N_{\rm MW}$, and the number of the DM particles originating from the LMC, $N_{\rm LMC}$, in the Solar region for the best fit Sun's position are listed in table~\ref{tab:Localrho} for the 15 MW-LMC analogues at the snapshot closest to LMC's first pericenter approach. While there are $[7,760-20,001]$  DM particles from the MW in the Solar region, the number of DM particles originating from the LMC in the Solar region is in the range of $[1 - 361]$. Due to this limited number of LMC particles in the Solar region, we are not sensitive
to the variation of the distribution of DM particles from the LMC within our defined Solar region. The low number of DM particles originating from the LMC is, therefore, a limitation of the current cosmological simulations as compared to idealized simulations, which can achieve a better resolution. Nevertheless, due to their high relative velocities with respect to the Sun, the DM particles from the LMC are more numerous compared to the high speed DM particles of the MW, and can significantly affect the high speed tails of the local DM velocity distribution  (as discussed below in section~\ref{sec: velocity distributions}). Therefore, the low value of $N_{\rm LMC}$ is not a major concern for the validity of our results.

\subsection{Local dark matter density}
\label{sec: DM density}

We first extract the local DM density, $\rho_\chi$, in the Solar region for the best fit Sun's position for the 15 MW-LMC analogue systems in Auriga at the snapshot closest to LMC's first pericenter approach. The results are given in table~\ref{tab:Localrho}. 
The local DM density is in the range of $\rho_\chi=[0.21-0.60]$~GeV/cm$^3$. This agrees  with the values obtained previously for the local DM density of MW-like halos in the EAGLE and APOSTLE~\cite{Bozorgnia:2016ogo}, and Auriga~\cite{Bozorgnia:2019mjk} simulations. It also agrees well with the local~\cite{Salucci:2010qr, Smith:2011fs, Bovy:2012tw, Garbari:2012ff, Zhang:2012rsb, Bovy:2013raa, 2018A&A...615A..99H, Buch:2018qdr} and global~\cite{McMillan:2011wd, Catena:2009mf, Weber:2009pt, Iocco:2011jz, Nesti:2013uwa, Sofue:2015xpa, Pato:2015dua, deSalas:2019pee} estimates from observations. The large range of local DM densities obtained from simulations is due to halo-to-halo variations and depends on halo properties such as mass (in our case the simulated halos have a mass to within less than a factor of 2 of that estimated for the MW halo~\cite{Callingham:2018vcf}), concentration, formation history, and mass of the stellar disk.

Next, we extract the percentage of the DM particles in the Solar region originating from the LMC analogue, $\kappa_{\rm LMC}$, at the snapshot closest to LMC's first pericenter approach. We consider a DM particle to have originated from the LMC analogue if it is bound to the LMC at infall as identified by the SUBFIND algorithm, and its distance from the center of the LMC at infall is less than twice the virial radius of the LMC at infall\footnote{The  majority of the  DM particles from the LMC in the Solar neighbourhood are typically found 
well within the virial radius of the LMC at infall and are therefore highly bound to the LMC at infall. Thus, our results are robust with respect to  the way we select the DM particles that have originated from the LMC analogue.}. $\kappa_{\rm LMC}$ is defined as the ratio of the number of DM particles originating from the LMC analogue in the Solar region and the total number of DM particles in the Solar region, multiplied by 100 to obtain the percentage. For the 15 MW-LMC analogues, $\kappa_{\rm LMC}$ in the Solar region for the best fit Sun's position is in the range of $[0.0077-2.8]$\%, as listed in table~\ref{tab:Localrho}. 
In the fourth column of the table,  we present the range that $\kappa_{\rm LMC}$ varies for each halo due to the different allowed Sun's positions.

\begin{figure}[t]
    \centering
    \includegraphics[width=\textwidth]{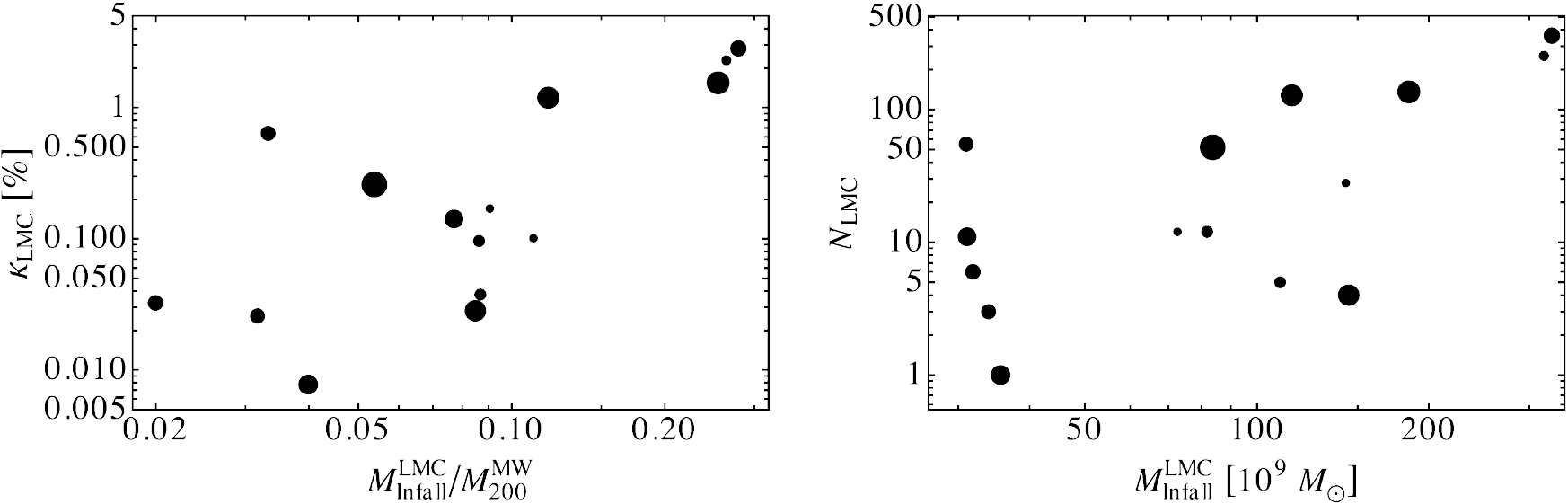}
    \caption{
    The correlation between $\kappa_{\rm LMC}$ and $M_{\rm Infall}^{\rm LMC}/M_{200}^{\rm MW}$ (left), and $N_{\rm LMC}$ and $M_{\rm Infall}^{\rm LMC}$ (right) for the 15 MW-LMC analogues. $\kappa_{\rm LMC}$ and $N_{\rm LMC}$ are given in the Solar region for the best fit Sun's position at the simulation snapshot closest to the LMC's pericenter approach. The sizes of points increase with the distance of the LMC analogues from host at pericenter.
    }
    \label{fig: Correlations}
\end{figure}

To investigate the reason for the halo-to-halo variation in $\kappa_{\rm LMC}$ and $N_{\rm LMC}$, in figure~\ref{fig: Correlations} we present the variation of these parameters with $M_{\rm Infall}^{\rm LMC}/M_{200}^{\rm MW}$ and $M_{\rm Infall}^{\rm LMC}$, respectively. The point sizes increase with the distance of the LMC analogues from host at pericenter. The left panel of the figure shows that in general, systems with a larger LMC to MW halo mass ratio also have a larger percentage of LMC particles in the Solar region in most cases. However, the two parameters are not tightly correlated. In particular, systems with similar $M_{\rm Infall}^{\rm LMC}/M_{200}^{\rm MW}$ can still show a large variation in $\kappa_{\rm LMC}$. This is mainly due to the variation in the distance of the LMC analogues from host at pericenter, $r_{\rm LMC}$, for these systems. A larger $r_{\rm LMC}$ translates to smaller $\kappa_{\rm LMC}$ for systems with similar LMC to MW mass ratio. Similarly, the right panel of the figure shows a degree of correlation between $N_{\rm LMC}$ and $M_{\rm Infall}^{\rm LMC}$, while there exists a degree of inverse correlation between $N_{\rm LMC}$ and $r_{\rm LMC}$ for systems with similar $M_{\rm Infall}^{\rm LMC}$.

We have also checked how $\rho_\chi$ and $\kappa_{\rm LMC}$ vary if we change the size of our defined Solar region. In particular, for the re-simulated halo 13 at the present day snapshot, decreasing the opening angle of the cone from $\pi/4$ to $\pi/6$ while keeping the spherical shell width the same, cuts $N_{\rm LMC}$ and $N_{\rm MW}$ by half, decreases  $\rho_\chi$ by $\sim 30\%$,  and increases $\kappa_{\rm LMC}$ by $\sim 20\%$, compared to the original Solar region. Decreasing the shell width from $6-10$~kpc to $7-9$~kpc while keeping the opening angle of the cone the same has a similar effect on $N_{\rm LMC}$ and $N_{\rm MW}$, but leads to an increase of $\sim 2\%$ in $\rho_\chi$  and $\sim 10\%$ in $\kappa_{\rm LMC}$. Decreasing both the opening angle of the cone to $\pi/6$ and the shell width to $7-9$~kpc, reduces $N_{\rm LMC}$ to $1/3$ and $N_{\rm MW}$ to $1/4$ of their original values, decreases $\rho_\chi$ by $\sim 25\%$,  and increases  $\kappa_{\rm LMC}$ by $\sim 35\%$. These changes are smaller than the halo-to-halo variation in these parameters, as it can be seen from table~\ref{tab:Localrho}.

\subsection{Dark matter velocity distributions}
\label{sec: velocity distributions}

Next we extract the DM speed distributions in the Solar region in the Galactic reference frame. For each halo, the  velocity vectors of the DM particles are specified with respect to the halo center. The normalized DM  speed distribution, $f(v)$, is given by
\be
f(v) = v^2 \int d\Omega_{\mathbf{v}}\Tilde{f}(\mathbf{v}) \; ,
\ee
where  $d\Omega_{\mathbf{v}}$ is an infinitesimal solid angle around the direction $\mathbf{v}$, and  $\Tilde{f}(\mathbf{v})$ is the normalized DM velocity distribution such that $\int dv f(v)=\int d^3v \Tilde{f}(\mathbf{v}) = 1$.

In the SHM, the local circular speed of the MW is usually set to 220~{\rm km/s}. To compare the local DM speed distributions of different halos, we scale the DM speeds in the Solar region for each halo by $(220~{\rm km/s})/v_c$, where $v_c$ is the local circular speed 
computed from the total mass enclosed within a sphere of radius 8~kpc for each halo. Moreover, we choose an optimal speed bin size of $25 \, \rm km/s$ to compute the DM speed distributions from the simulations. This bin size ensures that there are enough particles in each speed bin such that the statistical noise in the data points remains small, without smearing out any possible features in the DM speed distributions.

\begin{figure}[t]
    \centering
    \includegraphics[width=0.8\textwidth]{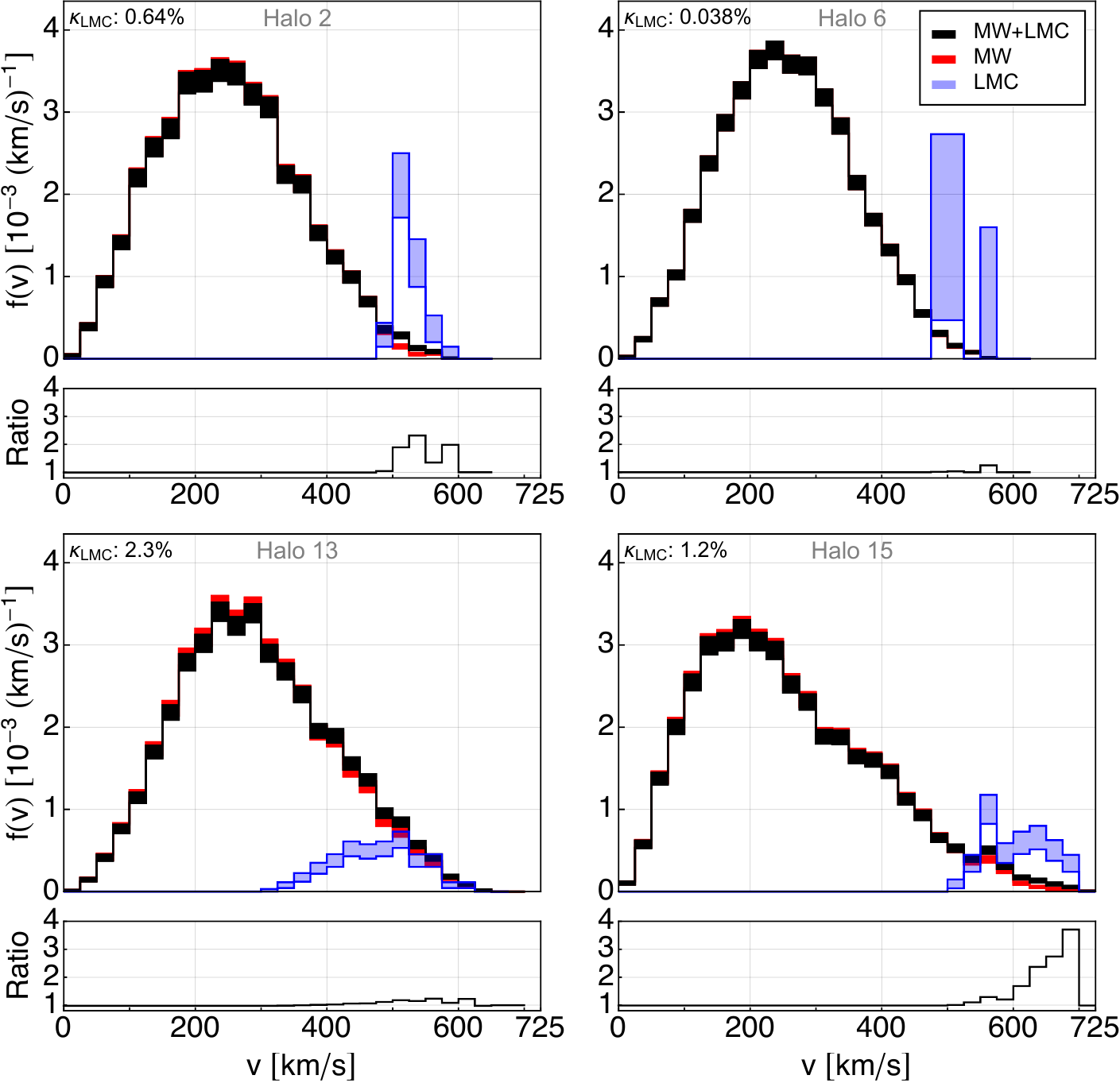}
    \caption{
    DM speed distributions in the Galactic rest frame in the Solar region for the best fit Sun's position  for four representative MW-LMC analogues: halo 2 (top left), halo 6 (top right), halo 13 (bottom left) and halo 15 (bottom right), for the  snapshot closest to the LMC's pericenter approach. The distributions of the DM particles originating from the MW+LMC, MW only, and LMC only are shown as black, red, and blue shaded bands specifying the $1\sigma$ Poisson errors, respectively. The LMC-only distribution has been scaled down by a factor of 10 for better visualization. The percentage of the DM particles originating from the LMC in the Solar region, $\kappa_{\rm LMC}$, is also specified on each panel in the upper left corner. The panels below the speed distribution plots show the ratio between the MW+LMC  and the MW-only distributions.
    }
    \label{fig: fv lmc candidates}
\end{figure}

In figure~\ref{fig: fv lmc candidates} we present the DM speed distributions in the Galactic rest frame for four MW-LMC analogues in the Solar region specified by their best fit Sun's position, for the  snapshot closest to the LMC's pericenter approach. The speed distribution of the total DM particles (native to the MW\footnote{Notice that the DM particles native to the MW in this simulation snapshot are under the influence of the LMC and are different from the DM particles belonging to an isolated MW.}  or originating from the LMC) in the Solar region is shown as black shaded bands (specifying the $1\sigma$ Poisson errors), while the distribution of the DM particles native to the MW is shown in red. The blue shaded bands show the speed distributions of the DM particles originating from the LMC in the Solar region, scaled down by a factor of 10 for better visualization. The speed distribution of the total DM particles and those native to the MW are both normalized to 1. The percentage of the DM particles in the Solar region originating from the LMC is also specified in the top left corner of each panel. The panels below the speed distribution plots
show the ratio of the speed distribution of the total DM particles and the MW-only distribution.

Among the 15 MW-LMC analogues, the four halos presented in figure~\ref{fig: fv lmc candidates} are representative of the differences seen in the local speed distributions of the DM particles originating from the MW only, the LMC only and the combined MW+LMC. Halo 2 (top left) has an intermediate percentage of DM particles originating from the LMC in the Solar region ($\kappa_{\rm LMC}=0.64$\%). It also has a sharply peaked speed distribution, leading to noticeable differences between the tails of the MW+LMC and MW-only speed distributions, with their ratio reaching values greater than 2 in the tail. Halo 6  (top right) is an example of a halo for which even a small fraction of DM particles in the Solar region originating from the LMC  ($\kappa_{\rm LMC}=0.038$\%) can lead to   differences in the tail of its DM speed distribution, as seen from the ratio plot. Halo 13 (bottom left) has a high  fraction of DM particles originating from the LMC ($\kappa_{\rm LMC}=2.3$\%) with a broad speed distribution, leading to  mild differences  between the MW+LMC and MW-only speed distributions  across a large range of speeds. The ratio of the two distributions reaches similar values in halo 6 and halo 13, despite halo 13 having a $\kappa_{\rm LMC}$ which is $\sim 60$ times larger than halo 6. Finally, halo 15 (bottom right) with $\kappa_{\rm LMC}=1.2$\%, 
shows a large variation  between the MW+LMC and MW-only speed distributions in the high speed tail, with their ratio approaching 4.

In general, the speed distribution of DM particles originating from the LMC is found to peak at the high speed tail ($\gtrsim 500$~km/s with respect to the Galactic center) of the speed distribution of DM particles originating from the MW. This leads to variations in the tail of the MW+LMC speed distribution as compared to the MW-only distribution, although the degree to which the distributions vary is subject to large halo-to-halo scatter. The particular shape and width of the LMC's speed distribution in the Solar region for each MW analogue can  affect the variations in the tail of the MW+LMC distribution. For example, halos with an even larger $\kappa_{\rm LMC}$ (as listed in table~\ref{tab:Localrho}), do not necessarily show significant differences in their $f(v)$ with and without the LMC particles.

\begin{figure}[t]
    \centering
    \includegraphics[width=0.8\textwidth]{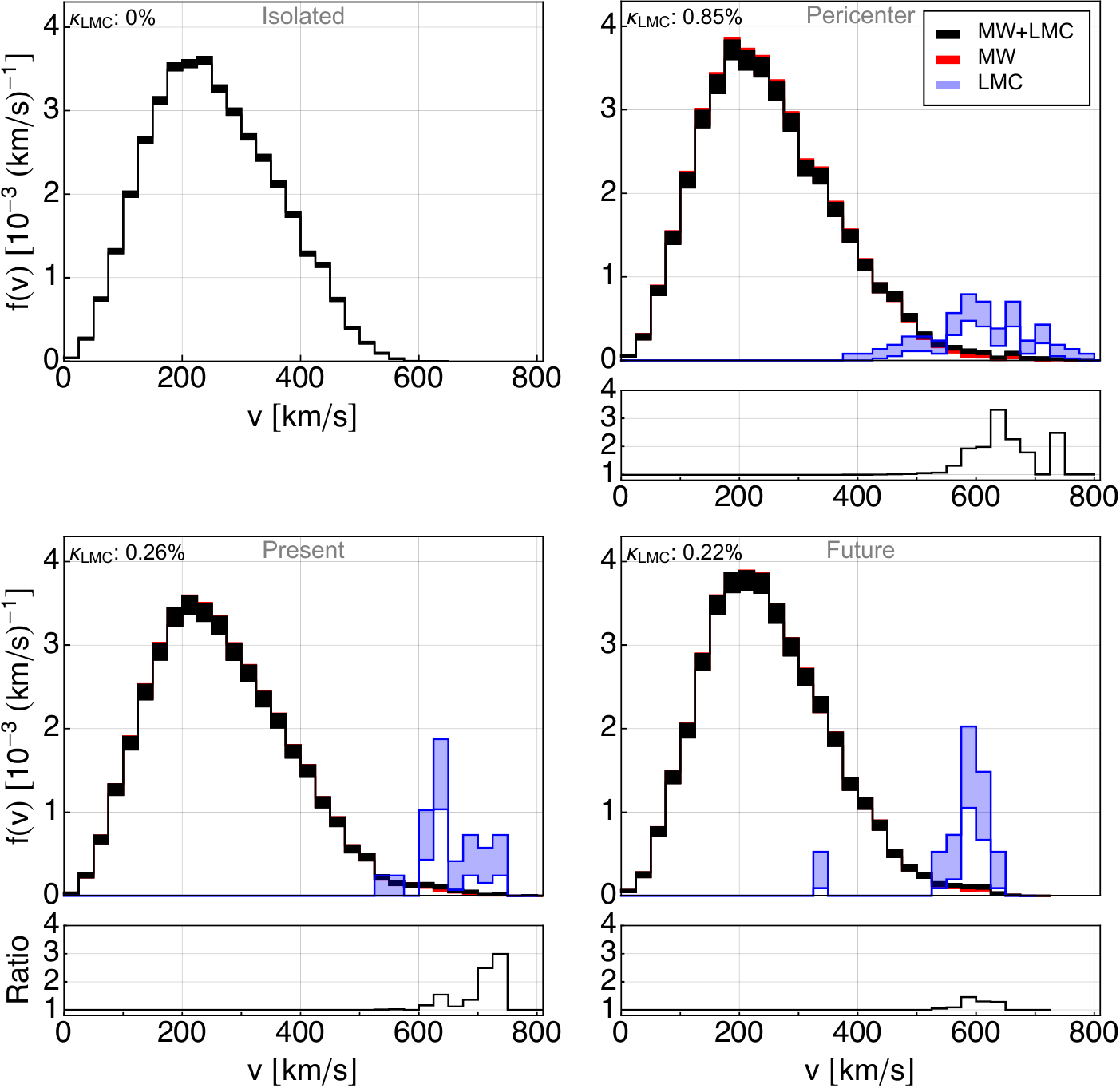}
    \caption{
    Local DM speed distribution in the Galactic rest frame for halo 13 for four representative snapshots: isolated MW (top left), LMC's pericenter approach (top right), present day MW-LMC (bottom left), and future MW-LMC (bottom right).
    The speed distributions of the DM particles originating from the MW+LMC, MW only, and LMC only are shown in black, red, and blue shaded bands representing the $1\sigma$ Poisson errors, respectively. The LMC-only distribution has been scaled down by a factor of 10. The distributions are presented in the Solar region for the best fit Sun's position, except for the isolated MW snapshot which is extracted in a spherical shell with radii between 6 and 10~kpc from the Galactic center. The value of $\kappa_{\rm LMC}$ is specified in the top left corner of each panel. The panels below each speed distribution plot show the ratio of the MW+LMC and the MW-only distributions, for all snapshots except the isolated MW. 
    }
    \label{fig: fv panel plot halo 49}
\end{figure}

In order to explore further the impact of the LMC on the local DM distribution during its orbit, we next focus on halo 13, where we rerun the simulations with finer snapshots close to the LMC's pericenter approach, as discussed in section~\ref{sec: selection criteria}. 
In figure \ref{fig: fv panel plot halo 49} we present the local DM speed distributions in the Galactic rest frame for halo 13 for the four snapshots representing different times in the LMC's orbit of the MW analogue (given in table~\ref{tab:snapshots}). The local speed distributions of the DM particles originating from the MW only (red), the LMC only (blue), and the MW+LMC (black) are shown. The distributions are presented in the Solar region for the best fit Sun's position for all snapshots 
except for the isolated MW, for which there is no LMC analogue and the best fit Sun's position cannot be defined. Hence, for the isolated MW the DM distribution is extracted in a spherical shell with radii between 6 to 10~kpc from the Galactic center. $\kappa_{\rm LMC}$ is also specified in each panel. The panels below the speed distribution plots show the ratio of the MW+LMC and the MW-only distributions, for all snapshots other than the isolated MW snapshot.

Figure~\ref{fig: fv panel plot halo 49} demonstrates that the LMC impacts the high speed tail of the local DM speed distribution, not only at its pericenter approach and at the present day, but also $\sim 175$~Myr after the present day. The value of $\kappa_{\rm LMC}$ is largest at pericenter and decreases as the LMC moves further from the host galaxy. Similarly, the ratio of the MW+LMC and the MW-only speed distributions in the high speed tail is largest at pericenter and decreases for the present day and future snapshots. For all  snapshots other than the isolated MW  (where $\kappa_{\rm LMC}=0$), the DM originating from the LMC has a speed distribution that peaks at the high speed tail of the native DM distribution of the MW, having a modest yet important impact on the total DM speed distribution. This is  similar to what we find in general for the 15 MW-LMC analogues at pericenter, as shown in figure~\ref{fig: fv lmc candidates}. 

Notice that when halo 13 is re-simulated with finer snapshots, the phase-space distribution of the DM particles  is not the same as in the original halo 13, as we are not comparing the two simulations at exactly the same time. As mentioned in section~\ref{sec: selection criteria}, the average time between snapshots is $\sim 150$~Myr in the original simulation, and it is difficult to precisely obtain the snapshot for the present day or LMC's pericenter approach. Furthermore, the Solar region for the best fit Sun's position is different in the original and the re-simulated halo 13, and this has a significant impact on the local DM velocity distribution. In particular, for the original halo 13, the cosine angles (eq.~\eqref{eq: cosine angles}) for the best fit Sun's position are $\cos \alpha=-0.796$ and $\cos\beta=-0.090$. Although these particular angles minimize the sum of the squared differences with their observed values, the value of $\cos\beta$ is very different than its observed value (as given in eq.~\eqref{eq: best fit cosine angles}), and it is therefore difficult to obtain a precise match to the observed Sun's position in the original halo 13. However, for the re-simulated halo 13, we obtain a much better match to the observed Sun's position (e.g.~$\cos \alpha=-0.995$ and $\cos\beta=-0.656$ for the best fit Sun's position at the present day snapshot). As a result, the speed distribution of DM particles from the LMC in the Solar region peaks at a noticeably higher speed in the re-simulated halo 13 compared to the original halo 13. Finally, there may also be a small variation in the phase-space distribution induced by the stochasticity of the baryonic physics model, which could lead to a slightly different evolution of the gravitational potential in the re-simulated halo. Hence, the local DM speed distributions and the values of  $\kappa_{\rm LMC}$  are also different between figures~\ref{fig: fv lmc candidates} and \ref{fig: fv panel plot halo 49} for halo 13. 

Our results in general confirm those presented in refs.~\cite{Donaldson:2021byu} and \cite{Besla:2019xbx}, which found that the small fraction of DM particles originating from the LMC in the Solar neighborhood (e.g.~$\sim 0.2\%$ in ref.~\cite{Besla:2019xbx}) dominates the high speed tail of the local DM speed distribution, in a suite of idealized simulations. Nevertheless, we note that the important effect of halo-to-halo variation in the results of our cosmological simulations cannot be overlooked.

\section{Halo integrals}
\label{sec: halo integrals}

The astrophysical dependence of the event rate in direct detection experiments (see section~\ref{sec: direct detection}) comes from the DM velocity distribution and density in the Solar neighbourhood. For the case of standard interactions, the \emph{halo integral} encodes the local DM velocity distribution dependence of the event rate  and is defined as
\begin{equation}
    \eta (v_{\rm min}, t) \equiv \int_{v>v_{\rm min}} d^{3}v\, \frac{\tilde{f}_{\rm det}(\mathbf{v},t)}{v} \; ,
    \label{eq: halo integral}
\end{equation}
where $\mathbf{v}$ is the relative velocity between the DM and the target nucleus or electron in the detector, with $v=|\mathbf{v}|$, $\tilde{f}_{\rm det}({\bf v},t)$ is the local DM velocity  distribution in the detector reference frame, and  $v_{\rm min}$ is the minimum  speed required for the DM particle to impart a recoil energy and momentum in the detector (given in eqs.~\eqref{eq: vmin} and \eqref{eq: vmin-e} for nuclear and electron recoils, respectively). Determining the influence of the LMC on the halo integrals in the Solar region directly reflects the expected change in direct detection event rates.

We extract the halo integrals of the MW-LMC analogues by boosting the local DM velocity distribution of each halo from the Galactic reference frame to the detector frame,
\begin{equation}
\tilde{f}_{\rm det}({\bf v},t) = \tilde{f}_{\rm gal}({\bf v}+{\bf v}_s+{\bf v}_e(t))\; , 
\end{equation}
where ${\bf v}_e(t)$ is the Earth's velocity with respect to the Sun, ${\bf v}_s={\bf v}_c +{\bf v}_{\rm pec}$ is the Sun's velocity in the Galactic rest frame, ${\bf v}_c$ is the Sun's circular velocity, and ${\bf v}_{\rm pec}=(11.10, 12.24, 7.25)$~km/s \cite{Schoenrich:2009bx} is the peculiar velocity of the Sun in Galactic coordinates with respect to the Local Standard of Rest. To boost the DM velocity distribution to the detector rest frame, we take $|{\bf v}_c|=v_c=220$~km/s. For simplicity, we neglect the small eccentricity of the Earth's orbit. In the following, we present the time-averaged halo integrals, which are averaged over one year.

\begin{figure}
\centering
\includegraphics[width=0.8\textwidth]{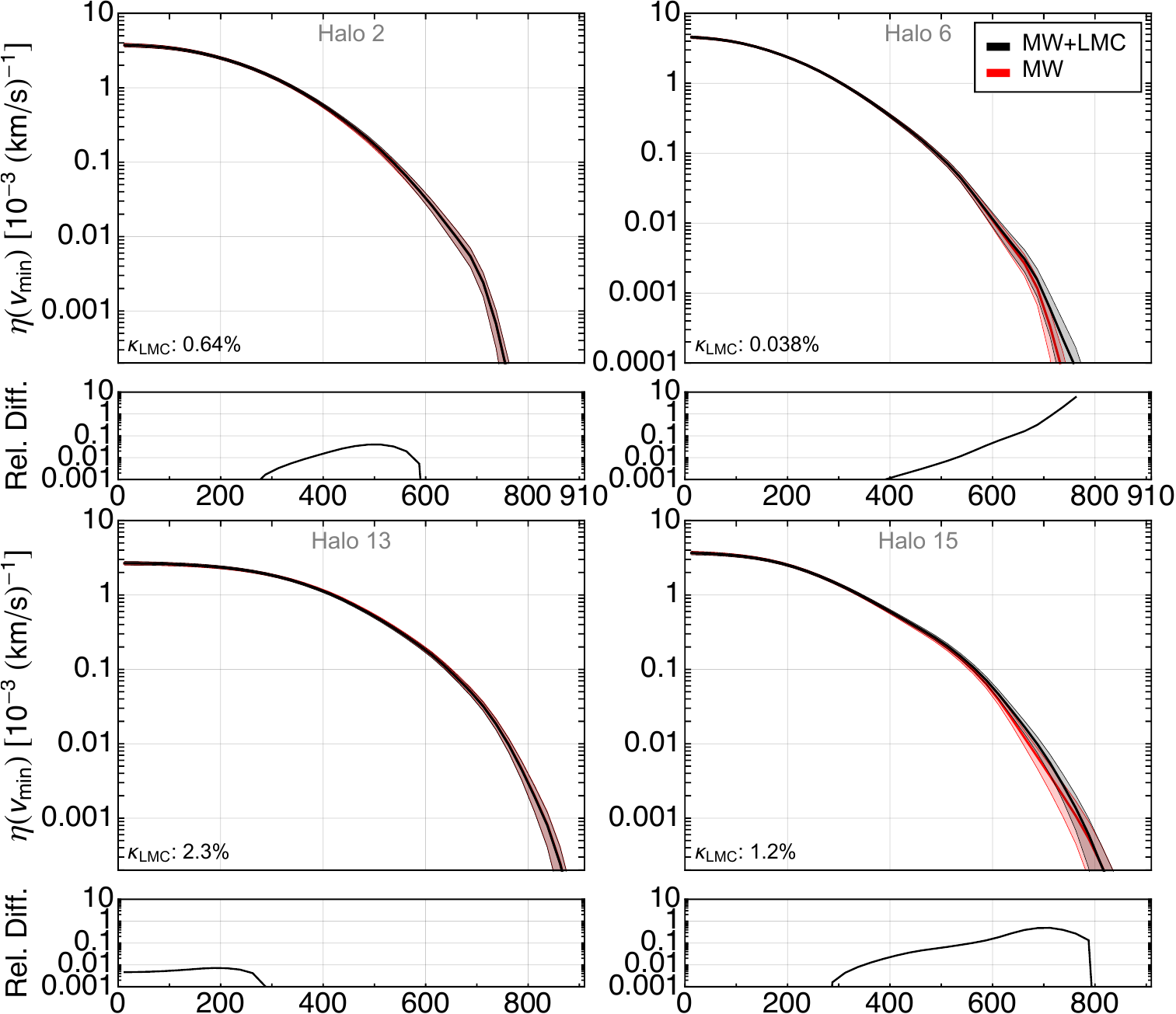}
 \caption{Time-averaged halo integrals for halos 2 (top left), 6 (top right), 13 (bottom left) and 15 (bottom right) in the Solar region for the best fit Sun's position, for the snapshot closest to the LMC's pericenter approach. The black and red curves show the halo integrals for the DM particles originating from the MW+LMC and MW only, respectively. In each case, the solid lines and the shaded bands correspond to the halo integrals obtained from the mean DM velocity distribution and the DM velocity distribution at $1\sigma$ from the mean, respectively. The value of $\kappa_{\rm LMC}$ is also specified on each panel. The panels below the halo integral plots show the relative difference between the MW+LMC and the MW-only halo integrals, $(\eta_{\rm MW+LMC}-\eta_{\rm MW})/\eta_{\rm MW}$.}
 \label{fig: halo panel lmc candidate}
\end{figure}

In figure~\ref{fig: halo panel lmc candidate} we present the time-averaged halo integrals as a function of $v_{\rm min}$  in the Solar region for the best fit Sun's position for the same four halos whose local DM speed distributions are shown in figure~\ref{fig: fv lmc candidates}: halos 2 (top left), 6 (top right), 13 (bottom left) and 15 (bottom right), for the
snapshot closest to the LMC's pericenter approach. The black and red solid lines are the halo integrals computed from the mean value of the velocity distributions of the DM particles originating from the MW+LMC and the MW only, respectively. The shaded bands correspond to the $1\sigma$ uncertainties in the halo integrals and are obtained from the DM velocity distribution at one standard deviation from the mean. The panels below the halo integral plots show the relative difference between the MW+LMC and the MW-only halo integrals, defined as $(\eta_{\rm MW+LMC}-\eta_{\rm MW})/\eta_{\rm MW}$.

As seen in figure~\ref{fig: halo panel lmc candidate}, halos 6 and 15 show some differences in the tails of the halo integrals of the MW+LMC and the MW-only, with their relative difference reaching $\sim 6$ for halo 6 and  $\sim 0.5$ for halo 15. The halo integrals of halos 2 and 13 do not show any visible deviations between the MW+LMC and the MW-only, and their relative differences are  smaller than 0.1 for halo 2 and 0.01 for halo 13. This is despite the fact that halo 13 has a higher $\kappa_{\rm LMC}$ in the Solar region compared to the other three halos. This highlights  the importance of the particular shape and peak speed of the LMC's speed distribution in the detector reference frame, in the Solar region of each MW analogue.

To quantify the changes in the tails of the halo integrals of the native DM particles of the MW and the total DM particles originating from MW+LMC, we define a dimensionless metric, 
\begin{equation}
\Delta \eta =\sum_{v_{\rm min}^i\geq 0.7 v_{\rm esc}^{\rm det}}\left[\eta_{\rm MW+LMC}(v_{\rm min}^i)-\eta_{\rm MW}(v_{\rm min}^i)\right]\Delta v_{\rm min}\; ,
    \label{eq: delta eta}
\end{equation}
where $\Delta v_{\rm min}$ is the bin size in $v_{\rm min}$, and $v_{\rm min}^i$ denotes the midpoint of the bins in $v_{\rm min}$ at which the halo integrals of the MW+LMC, $\eta_{\rm MW+LMC}$, and MW only, $\eta_{\rm MW}$, are evaluated. The sum runs over all bins with $v_{\rm min}^i$ larger than 70\% of the local escape speed from the MW in the detector rest frame, $v_{\rm esc}^{\rm det}$, which is estimated from the largest $v_{\rm min}$ where $\eta_{\rm MW}$ is nonzero. The values of $v_{\rm esc}^{\rm det}$ in the Solar region for the best fit Sun's position for the 15 MW-LMC analogues are given in table~\ref{tab:Localrho}, for the simulation snapshot closest to the LMC’s pericenter approach.

The metric in eq.~\eqref{eq: delta eta} reflects the changes in the exclusion limits set by direct detection experiments for the MW+LMC and MW-only distributions at low DM masses. Since the integral in eq.~\eqref{eq: halo integral} is computed for speeds greater than $v_{\rm min}$, and $v_{\rm min}$ depends inversely on the DM mass (eqs.~\eqref{eq: vmin} and \eqref{eq: vmin-e}), the exclusion limits in direct detection experiments become sensitive to small changes in the high speed tail of the halo integrals for low DM masses. Consequently, $\Delta\eta$ was defined to include only the differences in the halo integrals for $v_{\rm min}$ larger than 70\% of $v_{\rm esc}^{\rm det}$ to numerically reflect the variations in the tail of the halo integral and direct detection exclusion limits  at low DM masses. 
We have checked various other metrics for $\Delta\eta$, including the relative difference, the difference in the area under the curves, and considering different fractions of $v_{\rm esc}^{\rm det}$ in the above metric, with all showing similar general trends. The current definition 
preserves the global trends while providing the most intuitive connection between the halo integral plots and the direct detection exclusion limits (presented in section~\ref{sec: direct detection}) calculated therefrom.

In our analysis we find three key factors that contribute to changes in the tail of the halo integrals: 1) the percentage of DM particles originating from the LMC in the Solar region, 2) the Sun's position (and hence the \emph{Solar region}) in the simulations, and 3) the MW response due to the motion of the LMC as it traverses its orbit near pericenter. In the following sections we discuss how the results depend on each of these phenomena in detail.

%%%%%%%%%%%%%%%%%%%%%%%%%%%%%%%%%%%%%%%%%%%%%%%%%%%%%
\subsection{Impact of the DM particles originating from the LMC}
\label{sec: percent LMC}

\begin{figure}
     \centering
     \begin{subfigure}[b]{0.49\textwidth}
         \includegraphics[width=\textwidth]{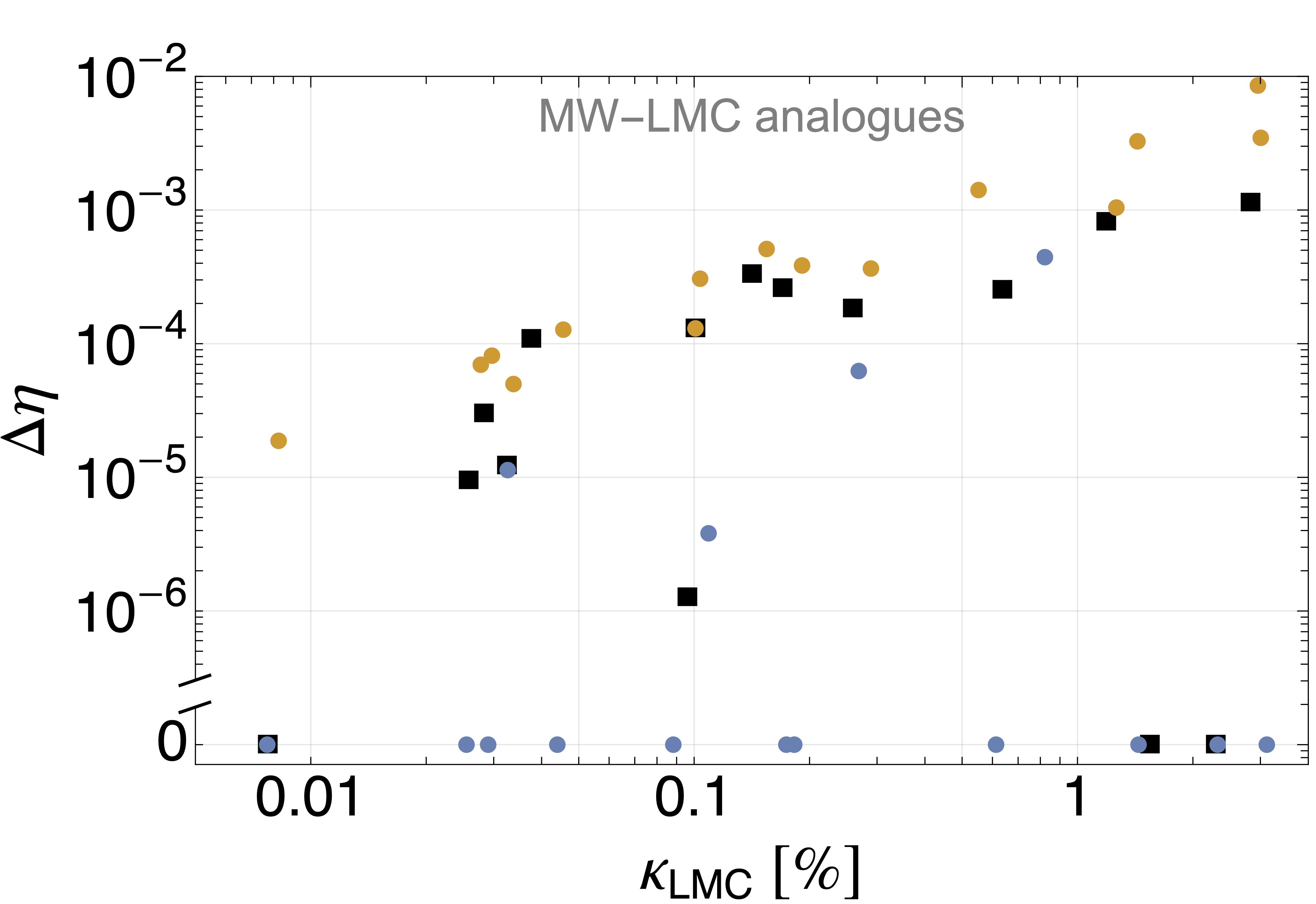}
     \end{subfigure}
    \hspace{-0.25cm}
     \begin{subfigure}[b]{0.49
     \textwidth}
         \includegraphics[width=\textwidth]{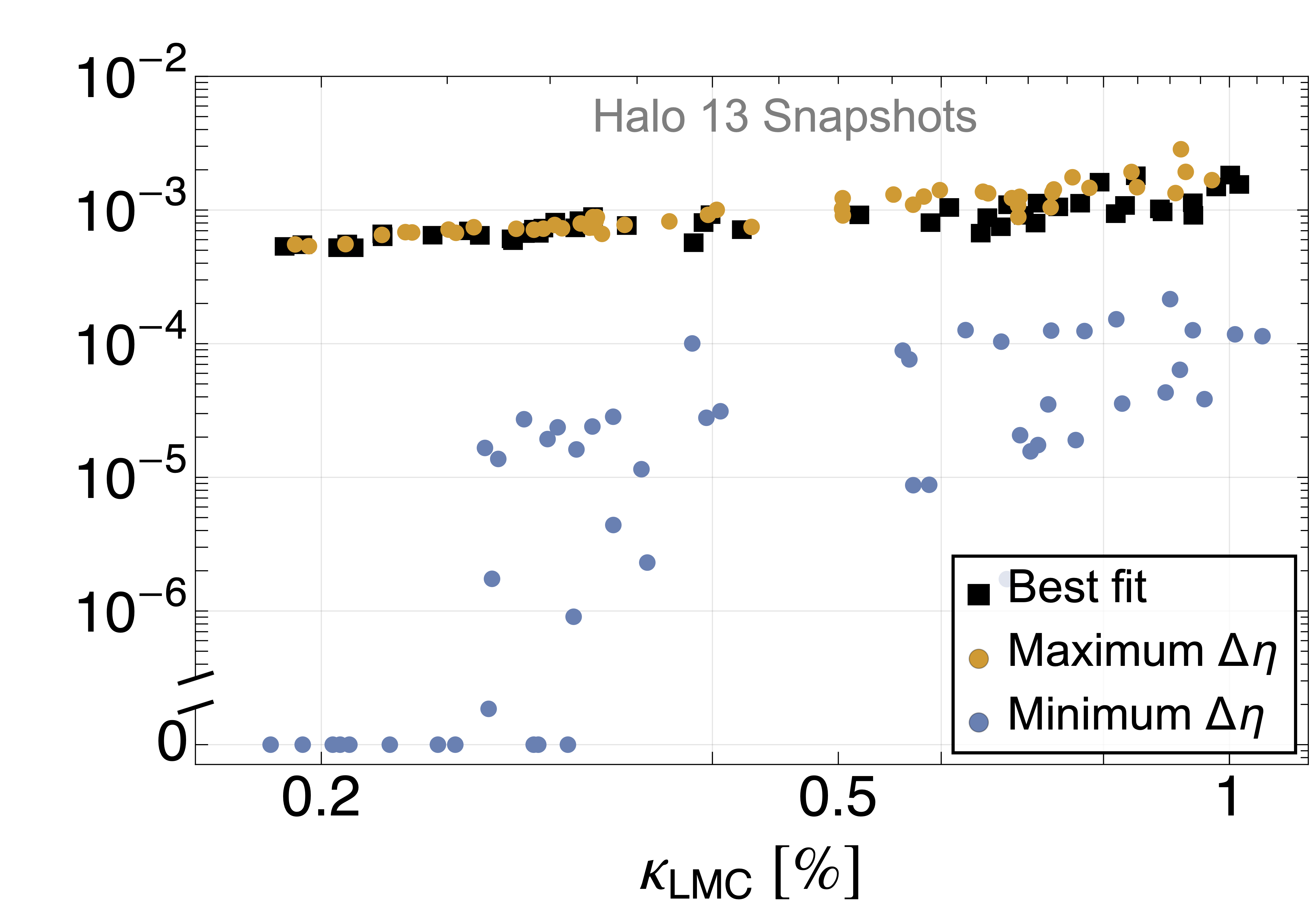}
     \end{subfigure}
        \caption{The correlation between the change in the tail of the halo integrals due to the LMC particles, $\Delta\eta$, and $\kappa_{\rm LMC}$ for three Solar regions: the Solar region for the best fit Sun's position (black squares), the Solar region that maximizes $\Delta\eta$ (yellow dots), and the Solar region that minimizes the $\Delta\eta$ (blue dots). Left panel: all 15 MW-LMC analogues at the snapshot closest to the LMC's pericenter approach. Right panel: different snapshots of halo 13, ranging from $\sim 314$~Myr before the present day to $\sim 175$~Myr after.}
        \label{fig: eta v percent}
\end{figure}

The DM particles originating from the LMC with high enough speeds become unbound to the LMC at infall, with some number at any given time ending up in the Solar region of the MW. Since these particles have on average higher speeds than the native DM particles of the  MW, it is expected that they will affect the high speed tail of the halo integrals, with a higher value of $\kappa_{\rm LMC}$ contributing to a more pronounced effect. 

Figure~\ref{fig: eta v percent} shows the correlation between the percentage of DM particles originating from the LMC in the Solar region, $\kappa_{\rm LMC}$, and the change in the tail of the halo integral due to the LMC particles, $\Delta\eta$ (eq.~\eqref{eq: delta eta}), for the best fit Sun's position (black squares), the Solar region that maximizes $\Delta\eta$ (yellow dots), and the Solar region that minimizes $\Delta\eta$ (blue dots). The left panel shows the results for the 15 MW-LMC analogues at the snapshot closest to the pericenter approach of the LMC, while the right panel is for different snapshots of halo 13, ranging from $\sim 314$~Myr before the present day to $\sim 175$~Myr after. In both panels, the $\Delta\eta$ for the best fit Sun's position is in general close to the maximum $\Delta \eta$, both showing an increase with $\kappa_{\rm LMC}$ in the Solar region. The minimum $\Delta \eta$ is zero or close to zero for a number of MW-LMC analogues and for some snapshots of halo 13, but still shows an increase with $\kappa_{\rm LMC}$ for halo 13. As discussed in section~\ref{sec: DM density}, $\kappa_{\rm LMC}$ generally increases with $M_{\rm Infall}^{\rm LMC}/M_{200}^{\rm MW}$. Therefore, a higher LMC to MW mass ratio would in general result in a larger $\Delta \eta$, depending on the particular Sun's position considered.

\begin{figure}[t]
    \centering
    \includegraphics[scale=.7]{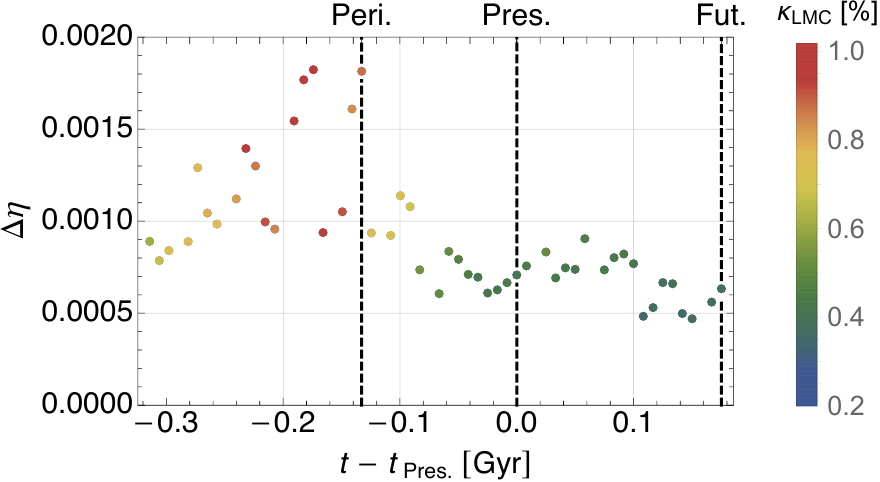}
    \caption{$\Delta \eta$ in the Solar region for the best fit Sun's positions for different snapshots in halo 13, plotted against the snapshot time relative to the present day snapshot, $t-t_{\rm pres}$. The snapshot times range from $\sim 314$~Myr before the present day to $\sim 175$~Myr after. The colour bar shows the range of $\kappa_{\rm LMC}$. The snapshots for the LMC's pericenter approach (Peri.), present day (Pres.) and $\sim 175$~Myr after the present day (Fut.) are specified with vertical dashed black lines.}
    \label{fig: bestfit deltaeta v snapshot}
\end{figure}

To better visualize the variation of $\Delta \eta$ in halo 13, in figure~\ref{fig: bestfit deltaeta v snapshot} we  present
$\Delta\eta$ in the Solar region for the best fit Sun's positions for different snapshots as a function of the snapshot time relative to the present day snapshot, $t-t_{\rm Pres.}$. The colour bar specifies the range of $\kappa_{\rm LMC}$. As expected,  snapshots with the highest value of $\kappa_{\rm LMC}$ near LMC's pericenter approach\footnote{The two snapshots occurring at $\sim 174$ and $\sim 191$~Myr before the present day snapshot are tied for the highest value of $\kappa_{\rm LMC}$ with 1.0\% each. From those, the one closest to the pericenter snapshot ($\sim 41$~Myr before LMC's pericenter approach) has the highest $\Delta\eta$.} also have the highest $\Delta\eta$, with snapshots far from pericenter dropping off in  both $\kappa_{\rm LMC}$ and $\Delta\eta$.

Notwithstanding the relationship between $\Delta\eta$ and $\kappa_{\rm LMC}$, there remains a scatter in the values of $\Delta\eta$ for systems with equal or similar values of $\kappa_{\rm LMC}$,  due to the particular choice of Sun's position for specifying the Solar region. This can be seen in both panels of figure~\ref{fig: eta v percent}, where there are large differences between the minimum and maximum $\Delta \eta$ for  the same or similar values of $\kappa_{\rm LMC}$. 
This leads us to consider not just the impact of $\kappa_{\rm LMC}$ on $\Delta \eta$, but also the effect of the exact Sun-LMC geometry, and whether the best fit Sun's position is a privileged position with respect to maximizing $\Delta\eta$. We explore this in the next section.

\subsection{Variation due to the Sun-LMC geometry}
\label{sec: Variation due to Sun-LMC Geometry}

For the 15 MW-LMC analogues we find a degree of variation in the values of $\kappa_{\rm LMC}$  due to the choice of the Solar region. In particular, $\kappa_{\rm LMC}$ can vary at most by a factor of $\sim 2$ depending on the MW-LMC analogue (see e.g.~the last column of table~\ref{tab:Localrho}). However, as discussed in section~\ref{sec: percent LMC}, for Solar regions with similar values of $\kappa_{\rm LMC}$ there is a large scatter in how much the tails of the MW+LMC halo integrals can deviate from their MW only counterparts.  This can also be seen in figure~\ref{fig: halo panel lmc candidate}, where the largest deviation in the tail of the halo integral is seen for halo 6 with $\kappa_{\rm LMC}=0.038\%$, while halo 13, with the highest $\kappa_{\rm LMC}$ of 2.3\%, shows a very small variation. 
This implies that the  value of $\kappa_{\rm LMC}$ is not the only important factor in specifying $\Delta \eta$, but  the particular Sun-LMC geometry of the chosen Solar region is similarly important.%

\begin{figure}
     \centering
         \includegraphics[width=0.85\textwidth]{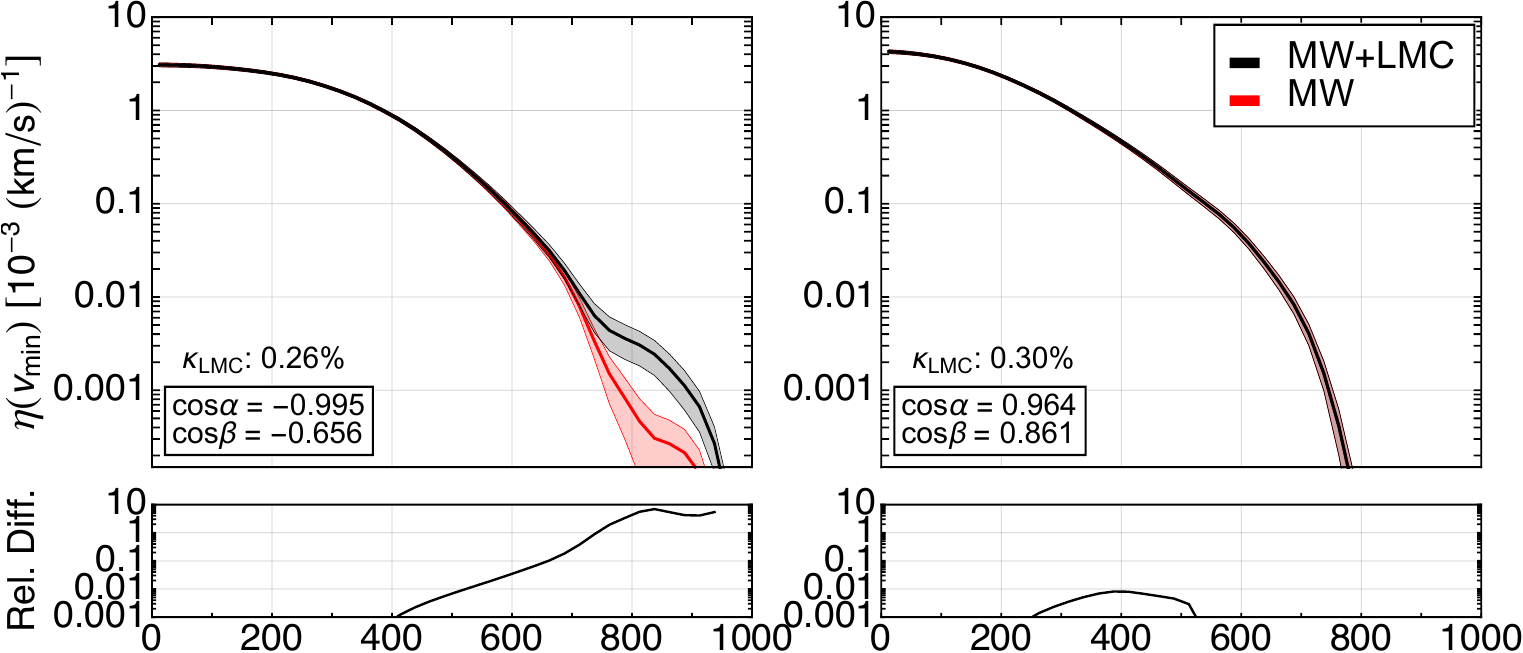}
        \caption{Time-averaged halo integrals for halo 13 at the present day snapshot for the MW+LMC (black) and the MW only (red) DM populations for the  best fit Sun's position (left panel) and the Solar region that has the minimum $\Delta\eta$ (right panel). The solid lines and the shaded bands correspond to the halo integrals obtained from the mean DM velocity distribution and the DM velocity distribution at $1\sigma$ from the mean, respectively. The value of $\kappa_{\rm LMC}$ and the cosine angles (eq.~\eqref{eq: cosine angles}) are also specified on each panel. The panels below the halo integral plots show the relative difference between the MW+LMC and the MW-only halo integrals.}
        \label{fig: halo panel bestfit}
\end{figure}

Figure~\ref{fig: halo panel bestfit} shows the time-averaged halo integrals for the MW+LMC (black) and the MW only (red) DM populations for the present day snapshot of halo 13 for two different Solar regions: the best fit Sun's position (left panel) and the Solar region that minimizes $\Delta\eta$ (right panel). The value of $\kappa_{\rm LMC}$ and the cosine angles corresponding to the particular Sun's position (eq.~\eqref{eq: cosine angles}) are also specified in each panel. The panels below the halo integral plots show the relative difference between the MW+LMC and the MW-only halo integrals. Clear differences between the tails of the MW+LMC and the MW halo integrals are visible in the case of the best fit Sun's position, while the variations in the halo integral tails are small for the Solar region that minimizes $\Delta \eta$. This figure highlights the importance of the Sun-LMC geometry, and that the same snapshot with similar values of $\kappa_{\rm LMC}$  can differ greatly in the tails of the halo integrals due to the choice of the Solar region.

\begin{figure}[t]
    \centering
    \includegraphics[width=\textwidth]{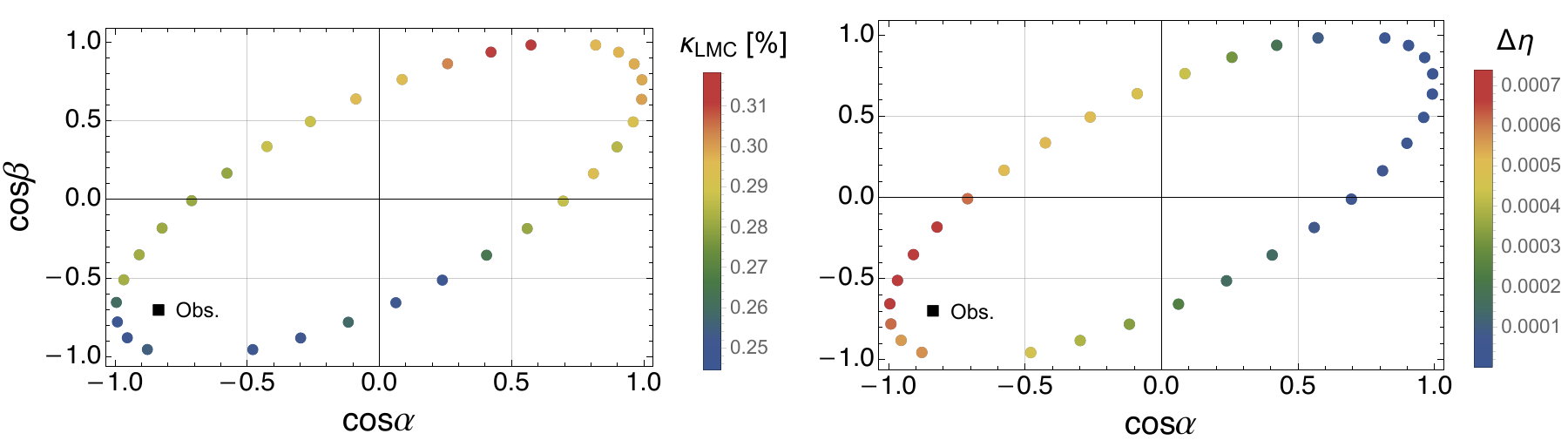}
    \caption{
    Cosine angles that parameterize the Sun-LMC geometry  (eq.~\eqref{eq: cosine angles})  for halo 13 at the present day snapshot for all allowed Sun's positions coloured by the value of $\kappa_{\rm LMC}$ (left panel) and $\Delta\eta$ (right panel). The observed values of the cosine angles (eq.~\eqref{eq: best fit cosine angles}) is specified with a black square on each panel.
  }
    \label{fig: cosine angles and sun position}
\end{figure}

To differentiate the effects on $\Delta \eta$ due to the choice of the Sun-LMC geometry from the effect of $\kappa_{\rm LMC}$, we study the cosine angles  that parameterize the Sun-LMC geometry, as given in eq.~\eqref{eq: cosine angles}. Figure~\ref{fig: cosine angles and sun position} shows the allowed Sun's positions in the parameter space of the two cosine angles for halo 13 at the present day snapshot. The colour bars show the range of $\kappa_{\rm LMC}$ and $\Delta\eta$ in the left and right panels, respectively. The black square in each panel shows the observed values of the cosine angles from eq.~\eqref{eq: best fit cosine angles}. Comparing the two panels of the figure shows that $\Delta\eta$ maximizes in the quadrant where $\cos \alpha$ and $\cos \beta$ are negative, despite $\kappa_{\rm LMC}$ reaching its maximum in the positive $\cos \alpha$ and $\cos \beta$ quadrant. 

The value of $\kappa_{\rm LMC}$ varies on average by $0.15\%$ between different allowed Sun's positions of a given snapshot, while it can vary by up to a percent between different snapshots. Hence, within a snapshot the dominant factor that impacts $\Delta\eta$ is the particular Sun-LMC geometry. Furthermore, we find that across snapshots $\Delta\eta$ tends to have its maximum values in the quadrants where  $\cos \beta$ is negative. A negative $\cos \beta$ indicates that the velocity vector of the LMC analogue is in the opposite direction of the Sun's velocity vector, leading to larger relative speeds of the particles originating from the LMC with respect to the Sun, and thus resulting in a larger $\Delta \eta$. Since  the observed cosine angles are also negative, this implies that variations in the tail of the halo integral for the best fit Sun's position should be close to the maximum possible variation from other allowed Sun's positions. This can also be seen from figure~\ref{fig: eta v percent}, where the values of $\Delta \eta$ for the best fit Sun's positions (black squares) are close to the maximum $\Delta \eta$ (yellow dots). We can therefore conclude that the best fit Sun's position is indeed in a privileged position with respect to maximizing $\Delta\eta$, by virtue of the observed $\cos \beta$ being negative, i.e.~the Sun's velocity and the LMC's velocity being predominantly in opposite directions. As a consequence, for the actual MW we expect the LMC to maximally affect the tail of the halo integral.

We have also checked the dependence of $\Delta \eta$  on the size of our defined Solar region for the present day snapshot of halo 13. We find that decreasing the size of the Solar region leads to a significant increase in $\Delta \eta$, due to better sensitivity to the best fit Sun's position. In particular, decreasing the opening angle of the cone from $\pi/4$ to $\pi/6$ while keeping the spherical shell width the same, increases $\Delta \eta$ by $18\%$. Decreasing the shell width to $7-9$~kpc while keeping the opening angle of the cone the same has a less significant effect and increases $\Delta \eta$ by $7\%$. Decreasing both the opening angle of the cone to $\pi/6$ and the shell width to $7-9$~kpc, increases $\Delta \eta$ by $78\%$. This comes at the cost of significantly reducing the number of DM particles from the MW and LMC in the Solar region (as discussed in section~\ref{sec: DM density}), and leading to very large Poisson uncertainties. Our results are therefore conservative with respect to the choice of the Solar region. Increasing the size of the Solar region, on the other hand, results in only a slight decrease in $\Delta \eta$, due to losing sensitivity to the best fit Sun's position. In particular, increasing the opening angle of the cone from  $\pi/4$ to $\pi/2$, while keeping the spherical shell width the same, decreases $\Delta \eta$ by only $2\%$.  

\subsection{MW response to the LMC}
\label{sec: MW response}

In addition to the particles originating from the LMC in the Solar region, the response of the local DM halo of the MW to the LMC's orbit can cause variations in the high speed tail of the local DM velocity distribution. The MW response to the LMC has been observed and studied before in idealized simulations~\cite{Besla:2019xbx, Donaldson:2021byu}, but it is important to test it in a fully cosmological setting where halos have multiple accretion events over their formation history. In this section, we explore the effect of this response on the tail of the halo integral in our cosmological simulations, and distinguish it from the effect of the high speed DM particles in the Solar region that originate from the LMC.

\begin{figure}[t]
    \centering
    \includegraphics[scale=0.55]{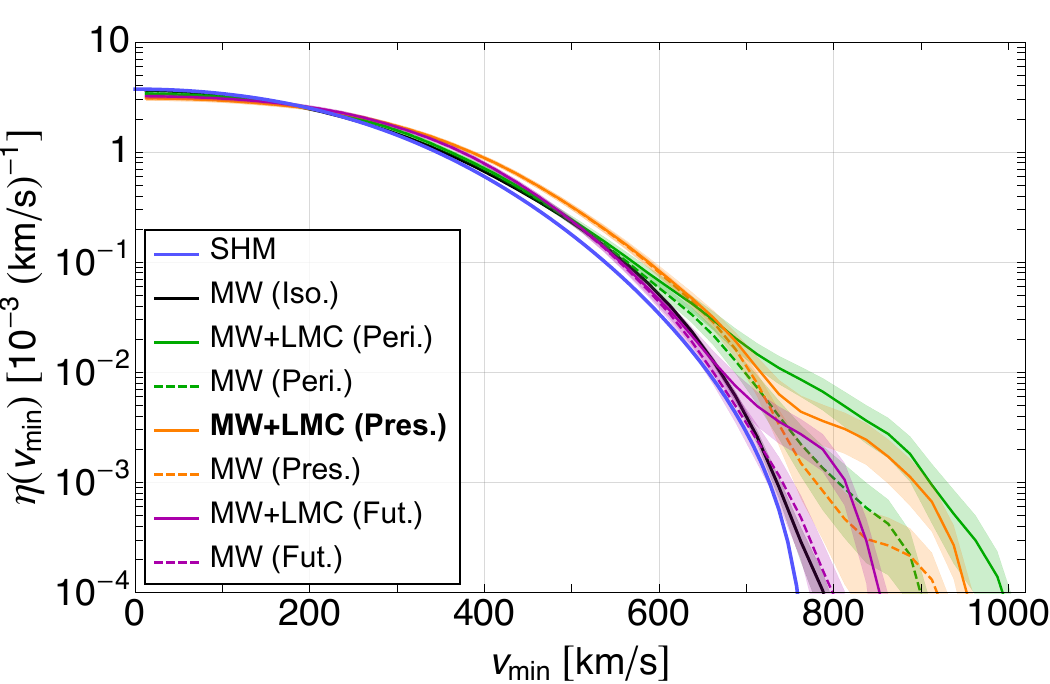}
    \caption{Time-averaged halo integrals for four  snapshots in halo 13: the isolated MW analogue (Iso.), the LMC's pericenter approach (Peri.), the present day MW-LMC analogue (Pres.), and the future MW-LMC analogue (Fut.). For each snapshot, the solid/dashed  lines and the shaded bands correspond to the halo integrals obtained from the mean DM velocity distribution and the DM velocity distribution at $1\sigma$ from the mean, respectively. For the present day, pericenter, and future snapshots the halo integrals are presented in the best fit Solar region
    for the MW+LMC (solid coloured curves) and MW-only (dashed coloured curves) DM populations. The isolated MW snapshot has no LMC analogue, so its MW halo integral (solid black curve) is shown for a Solar region defined as a spherical shell with radii between 6 and 10 kpc from the Galactic center. The solid blue curve shows the SHM halo integral obtained from a Maxwellian velocity distribution with a peak speed of 220~km/s and truncated at the escape speed of 544~km/s from the Galaxy.
   }
    \label{fig: halo int response}
\end{figure}

Figure~\ref{fig: halo int response} shows the time-averaged halo integrals for the four representative snapshots in halo 13: the isolated MW analogue (Iso.), the LMC's pericenter approach (Peri.), the present day MW-LMC analogue (Pres.), and the future MW-LMC analogue (Fut.). The halo integrals of the three latter snapshots are shown for the MW+LMC (solid coloured curves) and the MW only (dashed coloured curves) for the best fit Sun's position. The isolated MW snapshot has no LMC-like satellite and hence its halo integral (solid black curve) is extracted from the DM particles of the MW in a spherical shell with radii between 6 and 10 kpc from the Galactic center. For comparison, the blue curve shows the halo integral obtained from a Maxwellian velocity distribution with a peak speed of 220~km/s and truncated at the escape speed of 544~km/s from the Galaxy, as is commonly assumed in the SHM.

A comparison of the halo integral of the isolated MW  with  the MW-only halo integrals at the three other snapshots  shows how the native DM particles of the MW in the Solar region are boosted in response to the passage of the LMC. The isolated MW snapshot occurs $\sim 2.8$~Gyr before the present day snapshot and is a proxy for the  MW in the absence of the LMC's influence. The halo integral for this snapshot (solid black curve) is closest to the SHM halo integral, although its tail is slightly more extended, reaching $v_{\rm min} \sim 800~\rm{km/s}$. As the LMC reaches its first pericenter approach, the tail of the halo integral for the native DM population of the MW (dashed green curve) is boosted to $v_{\rm min} \sim 900~\rm{km/s}$. Since the present day LMC is not too far from its pericenter approach, the tail of the halo integral at the present day snapshot (dashed orange curve) shows a comparable boost to the pericenter snapshot. 
Finally, the tail of the halo integral for the local DM population of the MW returns to $v_{\rm min}  \sim 800$~km/s at the future MW-LMC snapshot (dashed magenta curve), which occurs $\sim 175$~Myr after the present day snapshot\footnote{This period of time is comparable to the overdensity wake induced by the passage of a satellite galaxy in a host DM halo in the Auriga simulations~\cite{Gomez2016}.}.

The addition of DM particles originating from the LMC in the Solar region further shifts the tails of the halo integrals to higher speeds.
In particular, the pericenter snapshot has the highest  $\kappa_{\rm LMC}$ of 0.85\% and also shows the highest boost in the tail of its MW+LMC halo integral (solid green curve), reaching $v_{\rm min} \sim 1000~\rm{km/s}$, which is $\sim100~\rm{km/s}$ higher than the reach of its MW-only counterpart. Similarly, with $\kappa_{\rm LMC}=0.26$\%, the present day snapshot has a  MW+LMC halo integral which exhibits a large difference compared its MW-only counterpart in the high speed tail. The future MW-LMC snapshot 
has $\kappa_{\rm LMC}=0.22$\%, and the tail of its  MW+LMC halo integral is boosted by $\sim50~\rm{km/s}$ compared to its MW-only counterpart. 

Comparing the boost of the native DM population of the MW in the pericenter and present day snapshots to the boost due to the presence of DM particles originating from the LMC reveals that the impact on the tail of the halo integral is of the same order of magnitude. 
Figure~\ref{fig: halo int response} demonstrates that the DM particles in the Solar neighborhood can be boosted from  $v_{\rm min} \sim 800~\rm{km/s}$ in the absence of the LMC (solid black curve) to more than $v_{\rm min} \sim 950~\rm{km/s}$ at the present day (solid orange curve), a combined increase of greater than $\sim 150~\rm{km/s}$ due to the MW response and the presence of high speed LMC particles in the Solar region.

\section{Implications for dark matter direct detection}
\label{sec: direct detection}

In this section we discuss the impact of the LMC on the interpretation of the results of DM direct detection experiments. In particular, in sections~\ref{sec: nuclear} and \ref{sec: electron} we consider the DM interaction with a target nucleus or electron, respectively, and study how the exclusion limits set by different direct detection experiments in the plane of the DM mass and scattering cross section change due to the presence of the LMC for a given experimental setup.

We simulate the signals in three different idealized direct detection experiments, which are inspired by near future detectors that would search for nuclear  or electron recoils due to the scattering with a DM particle. In order to find the constraints in the DM scattering cross section and mass plane, we employ the Poisson likelihood method  implemented in the {\rm DDCalc}~\cite{ddcalc} and {\rm QEDark}~\cite{QE, essig} software packages for nuclear and electron recoils, respectively. Taking the properties of the experiments and the local DM distribution as inputs, these packages provide the exclusion limits at a desired confidence level. To perform the direct detection analysis, we directly use the local DM velocity distributions extracted from the simulations.

\subsection{Dark matter - nucleus scattering}
\label{sec: nuclear}

In the case of DM-nucleus scattering, we consider a DM particle of mass $m_\chi$ scattering with a target nucleus of mass $m_T$ in an underground detector, and depositing the nuclear recoil energy, $E_R$. The differential events rate is given by
\begin{equation}
\frac{dR}{dE_R} = \frac{\rho_\chi}{m_\chi}\frac{1}{m_T} \int_{v>v_{\rm min}} d^3 v ~\frac{d\sigma_T}{d E_R}~v~\tilde{f}_{\rm det}({\bf v}, t)\; ,
\label{eq:difrate}
\end{equation}
where $\sigma_T$ is the DM-nucleus scattering cross section. 
Assuming elastic scattering, the minimum  speed required for a DM particle to deposit a recoil energy $E_R$ to the detector is given by
\begin{equation}
v_{\rm min}(E_R) = \sqrt{\frac{m_T E_R}{2 \mu^2_{\chi T}}}\; ,
\label{eq: vmin}
\end{equation}
where $\mu_{\chi T}$ is the reduced mass of the DM and nucleus.

For spin-independent interactions, the differential cross section is given by
\begin{equation}
  \frac{d\sigma_T}{dE_R}=\frac{m_T A^2 \sigma^{\rm SI}_{\chi N}}{2\mu_{\chi N}^2v^2}\,F^2(E_R)\; ,
\end{equation}
where $A$ is the atomic mass number of the target nucleus, $\sigma^{\rm SI}_{\chi N}$ is the spin-independent DM-nucleon scattering cross section at zero momentum transfer, $\mu_{\chi N}$ is the reduced mass of the DM and nucleon, and $F(E_R)$ is the spin-independent nuclear form factor for which we use the Helm form factor~\cite{Helm:1956zz}.

Hence, the differential event rate can be written in terms of the halo integral (eq.~\eqref{eq: halo integral}) as
\begin{equation}
    \frac{dR}{dE_R} = \frac{\rho_\chi A^2 \sigma^{\rm SI}_{\chi N}}{2 m_\chi \mu_{\chi N}^2}\,F^2(E_R)\,\eta (v_{\rm min}, t)\, .
\end{equation}

We consider two idealized direct detection experiments, one with a xenon target nucleus and the other with germanium. These detectors are based on the sensitivity of the  LZ~\cite{LZ:2022lsv, lz2015} and SuperCDMS~\cite{supercdms} direct detection experiments in the near future. Noble liquid detectors, such as LZ which has recently published its first results~\cite{LZ:2022lsv}, can reach large exposures and are sensitive to large DM masses and lower cross sections. On the other hand, cryogenic solid state detectors such as SuperCDMS are sensitive sub-GeV DM masses. Considered together, these two types of experiments probe a large range of DM masses and scattering cross sections.

For the xenon based experiment, we consider an energy range of $[2 -50]$~keV, an energy resolution of $\sigma_E=0.065 E_R+0.24\,{\rm keV}\sqrt{E_R/{\rm keV}}$~\cite{Bertone:2017adx}, and an exposure of $5.6\times 10^6$~kg~days with a maximum efficiency of 50\% as given in ref.~\cite{lz2015}. 
The exposure we consider for this experiment is expected to be achieved by LZ after five years of operation~\cite{lz2015}.

For the germanium based experiment, we consider two crystal target designs with different energy thresholds. The low energy threshold design is based on the projected high-voltage (HV) detector of the SuperCDMS SNOLAB experiment~\cite{supercdms}. We implement an energy  range of $[40 - 300]$~eV, with a constant signal efficiency of 85\%, a flat background level of 10 keV$^{-1}$~kg$^{-1}$~days$^{-1}$, and an exposure of $1.6\times10^4$~kg~days~\cite{supercdms, Kahlhoefer:2017ddj}. The high energy threshold design has similar features as the iZIP detector of the same experiment with a total exposure of $2.04\times10^4$~kg~days, an energy range of $[3 - 30]$~keV, 1 expected background event, and a flat efficiency of 75\%. The exposures considered are expected to be achieved by SuperCDMS after five years of operation~\cite{supercdms}.

\begin{figure}[t]
\centering
    \includegraphics[width=\linewidth]{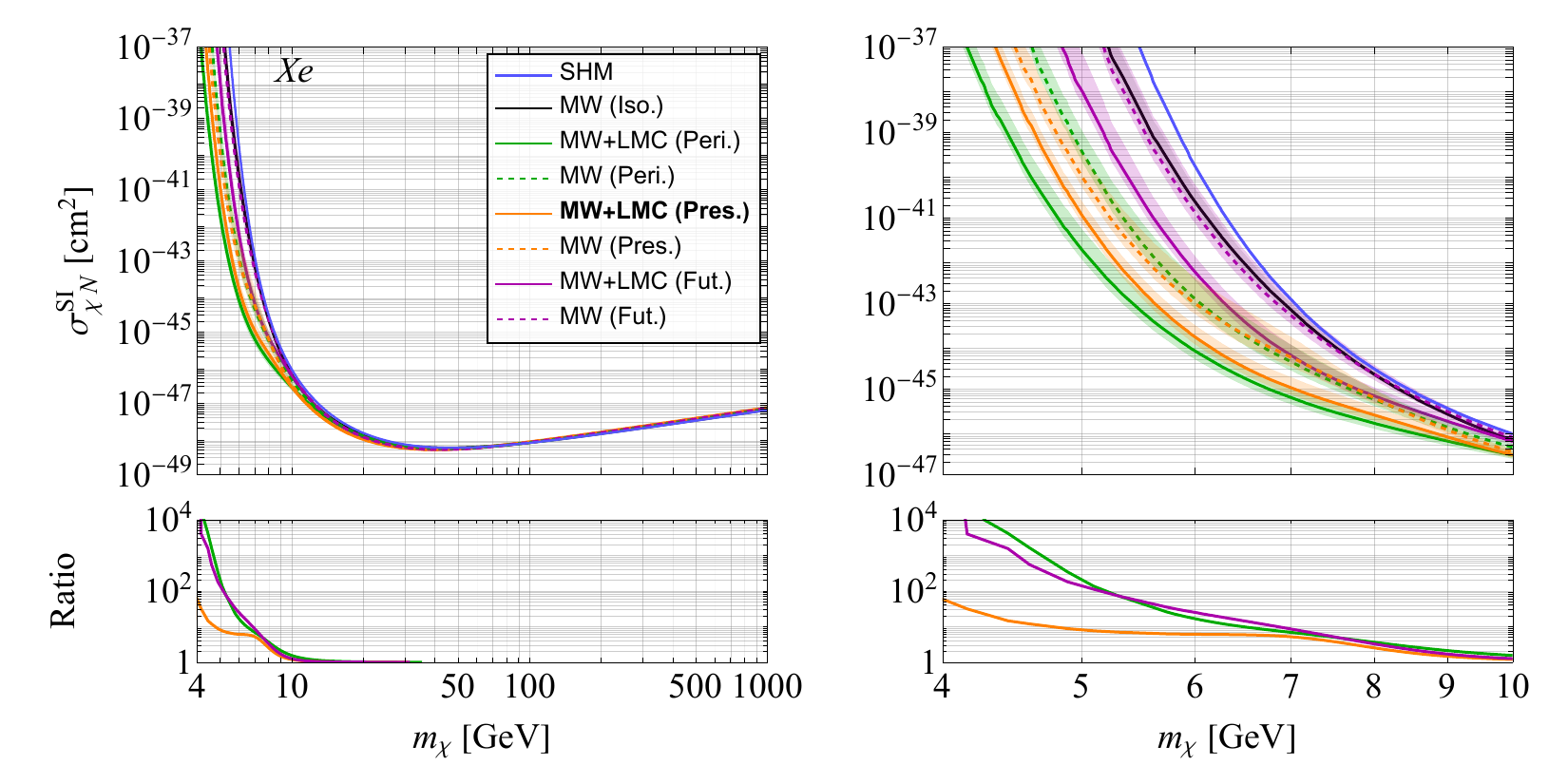}
    \caption{Top panels: exclusion limits at 90\% CL for a future xenon based experiment in the spin-independent DM-nucleon cross section and DM mass plane for four snapshots in halo 13: the isolated MW analogue (Iso.), the LMC's pericenter approach (Peri.),  the present day MW-LMC analogue (Pres.), and the future MW-LMC analogue (Fut.). For each snapshot, the solid/dashed lines and the shaded bands correspond to the exclusion limits obtained from the mean and the $1\sigma$ uncertainty band of the halo integrals, respectively. For the pericenter, present day, and future snapshots, the exclusion limits are presented in the Solar region for the best fit Sun's position for MW+LMC (solid coloured curves) and MW-only (dashed coloured curves) DM populations. For the isolated MW snapshot, the exclusion limit is shown for the DM population of the  MW  (solid black curve)  for a Solar region defined as a spherical shell with radii between 6 and 10 kpc from the Galactic center. The blue curve correspond to the exclusion limit for the SHM Maxwellian. The local DM density is set to $\rho_{\chi} = 0.3$~GeV/cm$^3$. Bottom panels: the ratios of the exclusion limits for the MW-only and the MW+LMC DM populations for the pericenter, present day, and future snapshots. The left panels show the limits and ratios for a large range of DM masses,
while the right panels zoom onto the low DM mass region.}
\label{fig:LZ_bestfit_merged}
\end{figure}

\begin{figure}[t]
\centering
\includegraphics[width=\textwidth]{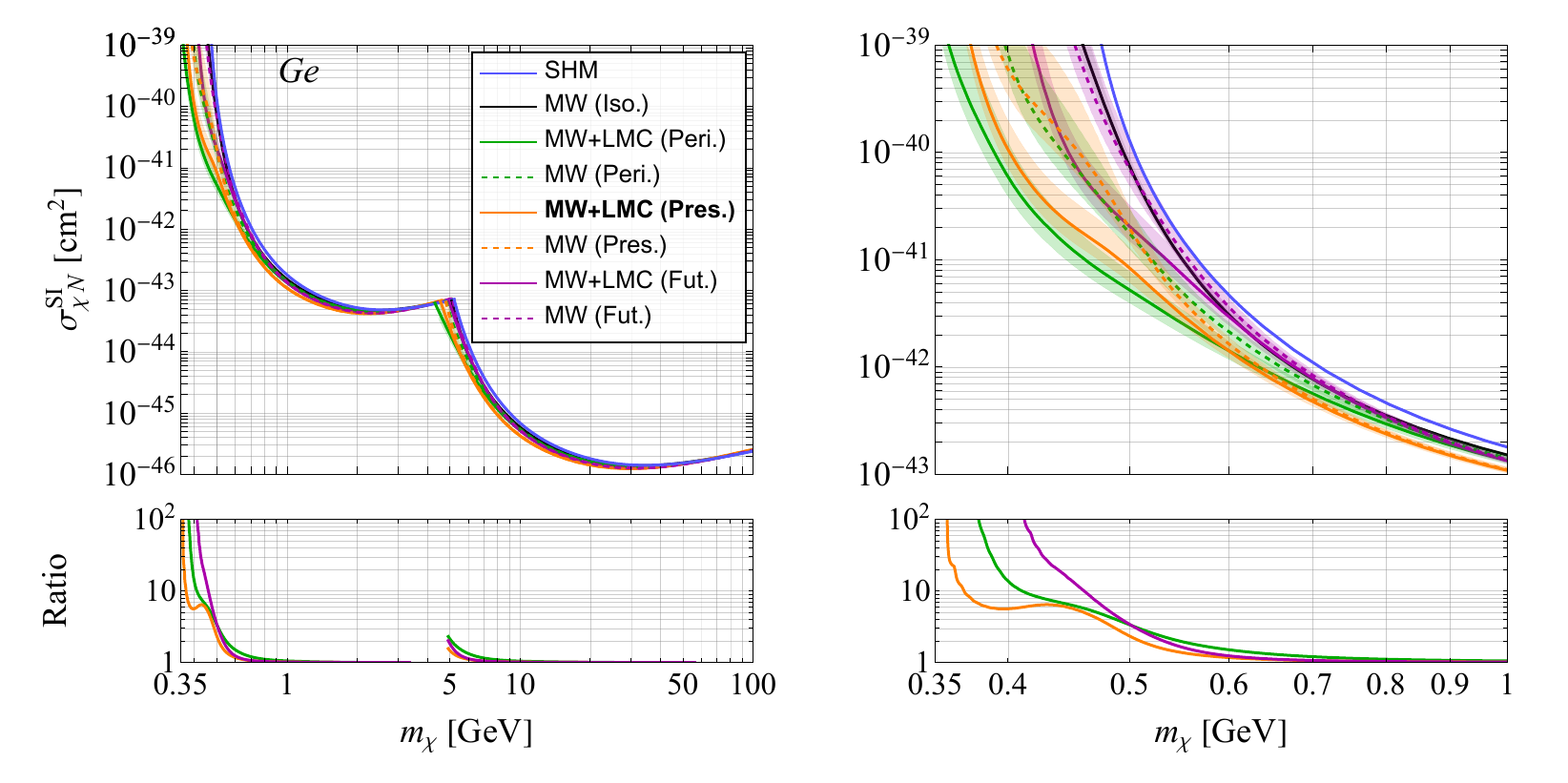}
\caption{Same as figure~\ref{fig:LZ_bestfit_merged}, but for a future germanium based experiment.}
\label{fig:supercdms_bestfit}
\end{figure}

The top panels of figures~\ref{fig:LZ_bestfit_merged} and \ref{fig:supercdms_bestfit}  show the exclusion limits at the 90\% CL in the plane of DM mass and spin-independent cross section set by the future xenon and germanium experiments using the local DM velocity distribution at the four representative  snapshots in halo 13, respectively. These snapshots are: the isolated MW analogue (Iso.), the LMC’s pericenter approach (Peri.),  the present day MW-LMC analogue (Pres.), and the future MW-LMC analogue (Fut.).  The mean and the shaded band in the exclusion limits are obtained from the mean and $1\sigma$ uncertainty band of the halo integrals shown in figure~\ref{fig: halo int response}, respectively. 
The exclusion limit for the isolated MW analogue is shown as the solid black curve, 
while the exclusion limits for the three other snapshots are shown as solid coloured curves for the MW+LMC distribution and dashed coloured curves for the MW-only distribution. For comparison, the exclusion limit assuming the SHM Maxwellian velocity distribution with peak speed of 220~km/s and truncated at the escape speed of 544~km/s from the Galaxy is shown as the solid blue curve. To distinguish the effect of the local DM velocity distribution, the local DM density is set to $\rho_\chi=0.3$~GeV/cm$^3$ in all cases\footnote{Since $\rho_\chi$ is a normalization in the event rate, changing its value results in the same upward or downward shift in all the exclusion limits.}, as is commonly adopted in the SHM.  The bottom panels of the figures show the ratios of the exclusion limits of the MW-only to the MW+LMC 
distribution  for  the pericenter, present day, and future snapshots. The left  panels show the limits and ratios for a large range of DM masses, while the  right panels zoom onto the low DM mass region to better visualize the differences at low masses.

The trends in figures~\ref{fig:LZ_bestfit_merged} and \ref{fig:supercdms_bestfit} are similar to those seen in figure~\ref{fig: halo int response} for the halo integrals of the different snapshots. In particular, the differences in the high speed tail of the halo integrals lead to variations in the exclusion limits at low DM masses, where the experiments are most sensitive to high values of $v_{\rm min}$. The isolated MW snapshot has the weakest exclusion limit at low DM masses and follows closely the SHM exclusion limit, while the DM distribution of the MW+LMC  at the LMC's pericenter approach leads to the strongest exclusion limit. As it can be seen from figure~\ref{fig:LZ_bestfit_merged}, for the xenon based experiment the exclusion limit for the MW+LMC distribution at the present day snapshot is lower than the isolated MW exclusion limit by an order of magnitude at $m_\chi \sim 8$~GeV, by more than three orders of magnitude  at $m_\chi \sim 6$~GeV, and by more than five orders of magnitude  at $m_\chi \sim 5$~GeV. Moreover at fixed cross sections, the exclusion limit for the MW+LMC distribution at the present day snapshot shows a shift of a few GeV towards smaller DM masses compared to the isolated MW for masses below $\mathcal{O}(10~{\rm GeV})$. Figure~\ref{fig:supercdms_bestfit} shows that for the germanium based experiment an order of magnitude  of vertical shift  occurs at $m_\chi \sim 0.5$~GeV between the exclusion limits of the MW+LMC distribution at the present day snapshot and the isolated MW, while the vertical shift is more than three orders of magnitude at  $m_\chi \sim 0.4$~GeV. Furthermore, at fixed cross sections and for DM masses below $\mathcal{O}(1~{\rm GeV})$, 
a horizontal shift of a few hundreds of MeV happens towards smaller DM masses. Hence, we can see from figures~\ref{fig:LZ_bestfit_merged} and \ref{fig:supercdms_bestfit} that the LMC extends the parameter space probed by direct detection experiments towards smaller DM masses.

Our results agree with those of ref.~\cite{Besla:2019xbx}, which also found that the presence of the LMC causes direct detection limits to shift to lower cross sections and lower DM masses, extending the sensitivity of those experiments. Hence, our results confirm that the findings of ref.~\cite{Besla:2019xbx} hold even in a fully cosmological setting.

\subsection{Dark matter - electron scattering}
\label{sec: electron}

In the case of DM-electron scattering, the differential event rate in a crystal target is given by~\cite{Essig:2015cda}
\begin{equation}
    \label{eq:diff_rate}
  \frac{dR}{ d\ln E_{e}  }  =  N_{\rm cell}\frac{\rho_\chi}{m_\chi}\frac{\overline\sigma_e \alpha m_e^2}{\mu_{\chi e}^2} \int d\ln q\, \frac{E_e}{q}\left[ |F_{\rm DM}(q)|^2 \,
              |f^{\rm crystal}(E_e, q)|^2 \,
                   \eta(v_{\rm min}(q, E_e)) \right]\; ,
\end{equation}
where $E_e$ is the energy deposited to the electron, $q$ is the  momentum transfer between the DM and the electron, $N_{\rm cell}$ is the number of unit cells per mass in the crystal target, $\overline\sigma_e$ is the DM-electron reference scattering cross section which parameterizes the strength of the interaction, $\alpha \simeq 1/137$ is the fine structure constant, $m_e$ is the mass of the electron, and $\mu_{\chi e}$ is the reduced DM-electron mass. The dimensionless crystal form factor, $f^{\rm crystal}$, encodes the dependence of the rate on the electronic structure of the target material. 

The DM form factor, $F_{\rm DM}$, gives the momentum dependence of the interaction. It can be shown that $F_{\rm DM}(q)=1$ for a contact interaction via a heavy mediator, $F_{\rm DM}(q)=(\alpha m_e/q)$ for an electron dipole moment coupling, and $F_{\rm DM}=(\alpha m_e/q)^2$ for a long-range interaction induced by the exchange of an ultralight or massless mediator~\cite{Essig:2015cda}.

Lastly, the minimum speed required for the DM particle in order for the electron to gain an energy $E_e$ with momentum transfer $q$ is given by
\begin{equation}
v_{\rm min}(E_e,q)=\frac{E_e}{q}+\frac{q}{2m_\chi}\, .
\label{eq: vmin-e}
\end{equation}

We consider a future silicon CCD experiment, based on the sensitivity of the next generation kg-sized DAMIC-M~\cite{Lee:2020abc, Castello-Mor:2020jhd, DAMIC-M:2023gxo} experiment. Direct detection experiments searching for DM-electron interactions provide a new avenue to probe MeV DM masses, due to the small mass of the electron. Semiconductors, in particular, have a very low ionization threshold of $\sim 1$~eV, and can be sensitive to single electron-hole pairs. We consider a silicon based detector with  an exposure of 1~kg~year and assuming zero background events, with an ionization threshold of 1 electron-hole pair. 

\begin{figure}[t]
  \centering
  \includegraphics[width=\textwidth]{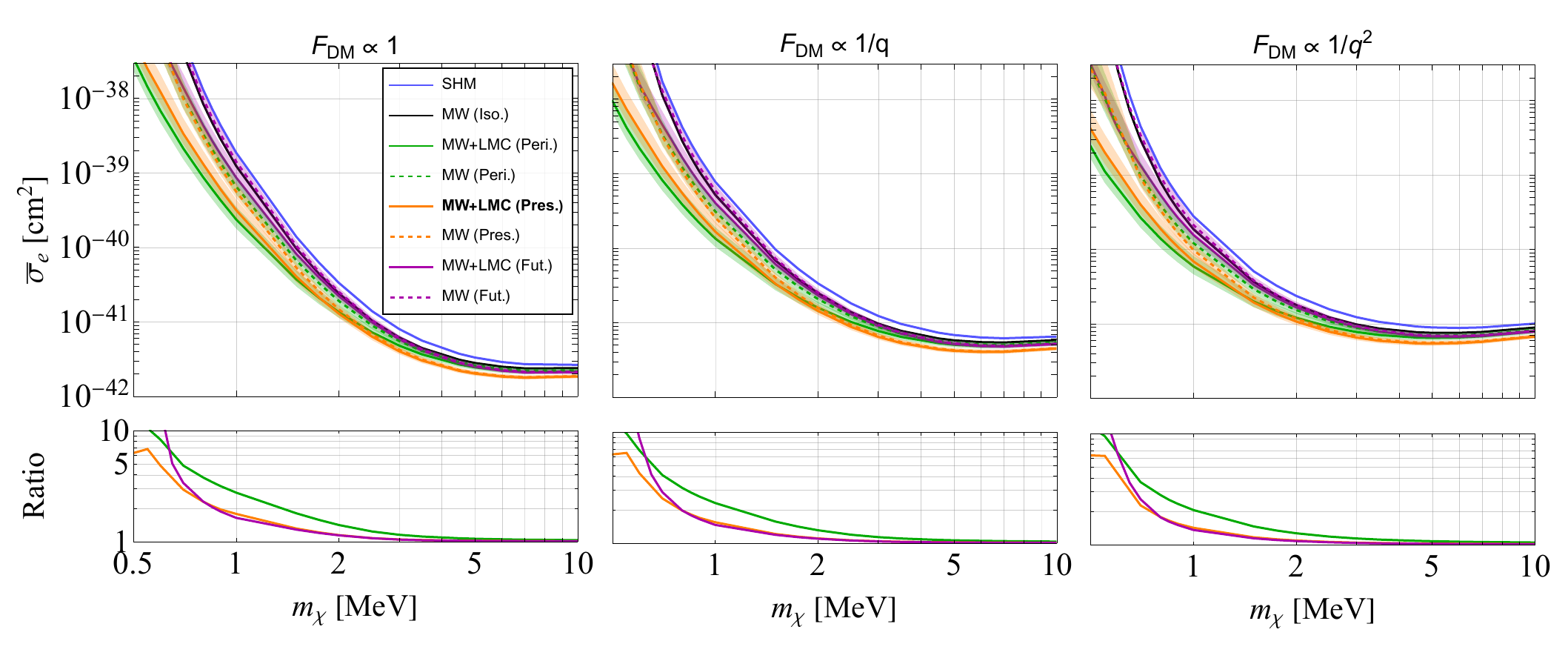}
  \caption{Top panels: exclusion limits at 95\% CL for a future silicon based experiment in the DM-electron cross section and DM mass plane for four snapshots in halo 13: the isolated MW analogue (Iso.), the LMC's pericenter approach (Peri.),  the present day MW-LMC analogue (Pres.), and the future MW-LMC (Fut.). The DM form factor is assumed to be $F_{\rm{DM}}=1$ (left panel), $F_{\rm{DM}} \propto q^{-1}$ (middle panel), and $F_{\rm{DM}} \propto q^{-2}$ (right panel). The blue curve corresponds to the SHM exclusion limit. The local DM density is set to $\rho_{\chi} = 0.3$~GeV/cm$^3$. Bottom panels: the ratios of 
  the exclusion limits for the MW-only and the MW+LMC  DM populations for the pericenter, present day, and future snapshots. The description of the different coloured curves are the same as in figure~\ref{fig:LZ_bestfit_merged}.
  }
  \label{fig:Si_electron_merged}
\end{figure}

The top panels of figure~\ref{fig:Si_electron_merged} show the exclusion limits at the 95\% CL in the plane of DM mass and DM-electron cross section for the future silicon based experiment, using the local DM velocity distribution at the isolated MW (black), pericenter (green), present day (orange), and future (magenta) snapshots of halo 13. The exclusion limits for the three latter snapshots are shown as
solid coloured curves for the MW+LMC distribution and dashed coloured curves for the MW-only distribution.  The mean and the
shaded band in the exclusion limits are obtained from the mean and $1\sigma$ uncertainty band of
the halo integrals shown in figure~\ref{fig: halo int response}, respectively. The SHM exclusion limit is shown as the solid blue curve. As for the DM-nucleus scattering, the local
DM density is set to $\rho_\chi = 0.3$~GeV/cm$^3$. The bottom panels show the ratio of the exclusion limits of the MW-only  to the MW+LMC distributions for the pericenter, present day, and future snapshots. The left, middle, and right panels show the results for three different DM  form factors, $F_{\rm{DM}}=1$, $F_{\rm{DM}} \propto q^{-1}$, and $F_{\rm{DM}} \propto q^{-2}$, respectively.

The general implications of the LMC for the exclusion limits on the DM-electron scattering cross section are similar to the DM-nucleus scattering, although the effect is smaller in the former case. As it can be seen from figure~\ref{fig:Si_electron_merged}, for all three choices of the DM form factor, the exclusion limits of the MW+LMC distribution at the LMC's pericenter approach and the present day MW-LMC show a shift towards smaller DM masses and lower DM-electron cross sections compared to the isolated MW. As expected, the shift becomes larger for smaller DM masses, where the experiment probes larger $v_{\rm min}$. In particular, the exclusion limit for the MW+LMC distribution at the present day snapshot is lower than the isolated MW exclusion limit by up to a factor $\sim 4$ at $m_\chi \sim 1$~MeV, and by up to a factor $\sim 50$ at $m_\chi \sim 0.6$~MeV. For DM masses below a few MeV and at fixed cross sections, the exclusion limit is shifted by a fraction of MeV towards smaller masses for all three choices of the DM form factor.

\section{Discussion and conclusions}
\label{sec: discussion}

In this work we have utilized a set of magneto-hydrodynamical simulations of MW-LMC analogues from the Auriga project~\cite{Grand:2016mgo} to study the effect of the LMC on the local DM  distribution and explore its implications for DM direct detection. We first identified  15 MW-LMC candidate systems by requiring that the stellar mass of the LMC analogue and its distance from the host at its first pericenter approach match observations. We then focused on one MW-LMC analogue and studied the impact of the LMC on the local DM distribution at different times (snapshots) in its orbit. In particular, we considered four representative snapshots: the isolated MW analogue, the first pericenter approach of the LMC analogue, the closest snapshot to the present day MW-LMC system, and the MW-LMC system at a future point in time, $\sim 175$~Myr after the present day.

We extracted the DM density and velocity distribution in the Solar region. 
The allowed positions of the Sun in the simulations were chosen such that they match the observed Sun-LMC geometry. In particular, we first found the stellar disk orientations in the simulations that make the same angle with the orbital plane of the LMC analogues as in observations. We then determined the position of the Sun for each allowed disk orientation by matching the angles between the orbital angular momentum of the LMC and the Sun's position and velocity vectors in the simulations to their observed values. The \emph{best fit Sun's position} was then defined as the one that leads to the closest match of the angles between the Sun's velocity vector and the LMC's position and velocities with observations. Using the local DM velocity distributions extracted from the simulations, we computed the halo integrals and showed how the LMC impacts their high speed tails. Finally, we simulated the signals in three near future xenon, germanium, and silicon direct detection experiments, considering the DM-nucleus interaction in the first two experiments and the DM-electron interactions in the latter, and studied the implications of the LMC on their exclusion limits. We summarize our findings below:

\begin{itemize}

\item The percentage of the DM particles originating from the LMC  in the Solar region is in the range of $[0.0077-2.8]$\% for the selected MW-LMC analogues. The local DM density is in the range of $[0.21-0.60]$~GeV/cm$^3$, depending on the halo.

\item The local speed distribution of the DM particles originating from the LMC peaks at the high speed tail ($\gtrsim 500$~km/s with respect to the  center of the MW analogue) of the local speed distribution of the native DM particles of the MW, with large halo-to-halo variations in the results. Focusing on different snapshots of one halo shows that the LMC impacts the high speed tail of the local DM speed distribution not only at its pericenter approach and the present day, but also up to $\sim 175$~Myr after the present day.

\item The LMC causes a shift in the high speed tail of the halo integrals towards larger speeds. Three key factors contribute to the variations in the tails of the halo integrals, quantified with the metric $\Delta \eta$ (eq.~\eqref{eq: delta eta}). First, a higher percentage of the DM particles originating from the LMC in the Solar region in general leads to a higher $\Delta \eta$, across different MW-LMC analogues and different snapshots of one system. Second, the exact Sun-LMC geometry for the choice of the Sun's position in the simulations 
has an impact on $\Delta \eta$, with the best fit Sun's position being close to a position which maximizes $\Delta \eta$. Therefore, in the MW we expect $\Delta \eta$ to be close to its maximum value at the Solar circle. Third, 
the native DM particles of the MW in the Solar region are boosted in response to the passage of the LMC, causing a further increase in $\Delta \eta$. The combination of  this boost and the presence of the high speed LMC particles in the Solar region causes a shift of greater than $\sim 150$~km/s in the high speed tail of the halo integrals at the present day.

\item The differences in the high speed tail of the halo integrals due to the LMC lead to considerable shifts in the expected direct detection exclusion limits  towards lower cross sections and smaller DM masses.  
In particular, the LMC lowers the exclusion limits set by the future xenon experiment on the DM-nucleon cross section by an order of magnitude for a DM mass of $\sim 8$~GeV, by more than three orders of magnitude for a DM mass of $\sim 6$~GeV, and by more than five orders of magnitude for a DM mass of $\sim 5$~GeV. For the future germanium experiment, the exclusion limits are lowered by an order of magnitude for a DM mass of $\sim 0.5$~GeV and by more than three orders of magnitude for a DM mass of $\sim 0.4$~GeV. The LMC also lowers the exclusion limits set by the future silicon experiment on the DM-electron cross section by up to a factor of $\sim 4$ for a DM mass of $\sim 1$~MeV and by up to a factor of $\sim 50$ for a DM mass of $\sim 0.6$~MeV. Furthermore, the LMC leads to a horizontal shift in the exclusion limits towards smaller DM masses, by a few GeV for xenon, a few hundred MeV for germanium, and a fraction of MeV for silicon, with the shift being more prominent for smaller DM masses. Thus, the LMC extends the parameter space probed by direct detection experiments towards lower DM masses.

\end{itemize}

The novel finding of our work is that the LMC's influence on the local DM distribution is significant even in a fully cosmological simulation, which follows the evolution of the MW and LMC analogues. While there are important halo-to-halo variations in  the results of our cosmological simulations,  a number of key conclusions could be reached by focusing on different snapshots of a particular MW-LMC analogue. Our study shows that a massive satellite that is just past its pericentric approach can significantly boost the high speed tail of the local DM velocity distribution. We also find that our particular Sun-LMC geometry maximizes the impact on the DM velocity distribution.

Our results are in general agreement with those of ref.~\cite{Besla:2019xbx}, which studied the effect of the LMC on direct detection signals in a suite  of idealized simulations of the LMC's orbit around the MW. Similar to our findings, they found that for small DM masses the LMC causes a vertical shift of more than an order of magnitude in the exclusion limits on the DM-nucleon cross section towards smaller cross sections.

The results of our fully cosmological simulations provide further evidence of the importance of the LMC's impact on the local DM distribution. It also strengthens the argument that these significant effects should not be overlooked in the analysis of future DM direct detection data, especially for low DM masses. Finally, our results have wider implications for the validity of utilizing the idealized simulations to understand other phenomena, such as the predictions for a DM wake induced by the LMC in the halo. Future cosmological simulations, which can achieve higher resolution would ultimately be able to quantify with high precision the differences in the high speed tail of the local DM velocity distribution due to the presence of the LMC.

%*******************************%

\acknowledgments

We thank Gianfranco Bertone for discussions on the results of this work. AS, NR, and NB acknowledge the support of the Natural Sciences and Engineering Research Council of Canada (NSERC), funding reference number RGPIN-2020-07138, and the NSERC Discovery Launch Supplement, DGECR-2020-00231. AF is supported by a UKRI Future Leaders Fellowship (grant no MR/T042362/1). GB acknowledges support from NSF CAREER award AST-1941096. CSF acknowledges support from the European Research Council (ERC)
Advanced Investigator grant DMIDAS (GA 786910). FAG acknowledges support from ANID FONDECYT Regular 1211370 and by the ANID BASAL project FB210003. FAG acknowledges funding from the Max Planck Society through a ``Partner Group'' grant. RG acknowledges financial support from the Spanish Ministry of Science and Innovation (MICINN) through the Spanish State Research Agency, under the Severo Ochoa Program 2020-2023 (CEX2019-000920-S). This work used the DiRAC Memory Intensive system at Durham University, operated by ICC on behalf of the STFC DiRAC HPC Facility (www.dirac.ac.uk). This equipment was funded by BIS National E-infrastructure capital grant ST/K00042X/1, STFC capital grant ST/H008519/1, and STFC DiRAC Operations grant ST/K003267/1 and Durham University. DiRAC is part of the National E-Infrastructure.

\clearpage

% BIBLIOGRAPHY
\bibliographystyle{JHEP}
\bibliography{refs}

\end{document}